\documentclass[12pt,preprint]{aastex}

\usepackage{amsmath}

\shorttitle{N-IR Polarimetry of Serpens}
\shortauthors{Sugitani et al.}

\begin{document}


\title{Near-Infrared Imaging Polarimetry of the Serpens Cloud Core:
Magnetic Field Structure, Outflows, and Inflows in A Cluster Forming Clump}


\author{Koji Sugitani,\altaffilmark{1} Fumitaka Nakamura,\altaffilmark{2} 
Motohide Tamura,\altaffilmark{3} Makoto Watanabe,\altaffilmark{4} 
Ryo Kandori,\altaffilmark{3} Shogo Nishiyama,\altaffilmark{5} 
Nobuhiko Kusakabe,\altaffilmark{3} Jun Hashimoto,\altaffilmark{3} 
Tetsuya Nagata,\altaffilmark{5 } and Shuji Sato\altaffilmark{6 }
}

\altaffiltext{1}{Graduate School of Natural Sciences, Nagoya City University, Mizuho-ku,
 Nagoya 467-8501, Japan; sugitani@nsc.nagoya-cu.ac.jp}
 \altaffiltext{2}{Faculty of Education and Human Sciences, Niigata University, 
 Niigata 950- 2181, Japan; fnakamur@ed.niigata-u.ac.jp.}
\altaffiltext{3}{National Astronomical Observatory, 2-21-1 Osawa, Mitaka, 
Tokyo 181-8588, Japan; motohide.tamura@nao.ac.jp, 
r.kandori@nao.ac.jp, nb.kusakabe@nao.ac.jp, Jun.Hashimoto@nao.ac.jp}
\altaffiltext{4}{Department of Cosmosciences, Hokkaido University, Kita 10, Nishi 8, Kita-ku, 
Sapporo, Hokkaido 060-0810, Japan; mwata@ep.sci.hokudai.ac.jp}
\altaffiltext{5}{Department of Astronomy, Kyoto University, Sakyo-ku, Kyoto 606-8502, 
Japan; shogo@kusastro.kyoto-u.ac.jp, nagata@kusastro.kyoto-u.ac.jp}
\altaffiltext{6}{Department of Astrophysics, Nagoya University, Chikusa-ku, 
Nagoya 464-8602, Japan; ssato@z.phys.nagoya-u.ac.jp}


\begin{abstract}
We made deep near-infrared ($JHK$s) imaging polarimetry toward the Serpens cloud core, 
which is a nearby, active cluster forming region.
The polarization vector maps show that the near-infrared reflection light 
in this region mainly originates from SVS2 and SVS20, 
and enable us to detect 24 small infrared reflection nebulae associated with YSOs.  
Polarization measurements of near-infrared point sources indicate 
an hourglass-shaped magnetic field, of which symmetry axis is nearly perpendicular to 
the elongation of the C$^{18}$O ($J=1-0$)  or submillimeter continuum emission. 
The bright part of C$^{18}$O ($J=1-0$),  submillimeter continuum cores as well as 
many class 0/I objects are located just toward the constriction region of 
the hourglass-shaped magnetic field.
Applying the Chandrasekhar \& Fermi method and taking into account the recent study 
on the signal integration effect for the dispersion component of the magnetic field, 
the magnetic field strength was estimated to be $\sim$100 $\mu$G, 
suggesting that the ambient region of the Serpens cloud core is moderately magnetically supercritical.
These suggest that the Serpens cloud core first contracted along the magnetic field to be an elongated cloud, 
which is perpendicular to the magnetic field, and that then the central part contracted 
cross the magnetic field due to the high density in the central region of the cloud core, 
where star formation is actively continuing. 

Comparison of this magnetic field with the previous observations of molecular gas and 
large-scale outflows suggests a possibility that the cloud dynamics is controlled by the magnetic field, 
protostellar outflows and gravitational inflows.
Furthermore, the outflow energy injection rate appears to be larger than 
the dissipation rate of the turbulent energy in this cloud, 
indicating that the outflows are the main source of turbulence  
and that the magnetic field plays an important role both in 
allowing the outflow energy to escape from the central region of the cloud core 
and enabling the gravitational inflows from the ambient region to the central region.
These characteristics appear to be in good agreement with the outflow-driven turbulence 
model and imply the importance of the magnetic field to continuous star formation 
in the center region of the cluster forming region.

\end{abstract}





\section{Introduction}

Stars are formed by gravitation in molecular clouds having both turbulence and  magnetic fields 
in the Galaxy,  and most of stars are thought to be formed in clusters \citep[e.g.,][]{lad03, al07}. 
A mass spectrum of prestellar condensations  is reported to have the power similar to 
that of the stellar IMF both in dust continuum observations \citep[][and references therein]{re06} 
and molecular-line observations \citep[e.g.,][]{ik07},   
and theoretical studies of turbulent molecular clouds \citep[][and subsequent works]{kl98} 
suggest that these condensations were formed through turbulent shock. 
One of  the most promising sources of ordinary turbulence is outflows from protostars,  
which are ubiquitous in star forming regions and are believed  to be formed 
through the mediation of magnetic field. 
Magnetic fields are also considered to play an important role in dynamical evolution of molecular clouds
and control of star formation, i.e., formation of molecular cloud cores and their collapse \citep[e.g.,][]{mck07}.

Recently, \cite{li06} and \cite{nakam07} presented realistic 3D MHD simulations of cluster formation, 
taking into account the effect of protostellar outflows as well as initial turbulence and a magnetic field. 
In their simulations, they indicated that the initial turbulence is quickly replaced by turbulence
generated by protostellar outflows, keeping the quasi-equilibrium state with a slow rate  of star formation, 
and that magnetic fields are dynamically important if their initial strengths are not far below 
the critical value for static cloud support because of  the amplification by the outflow-driven turbulent motions. 
The magnetic field is expected to influence the directions of outflow ejection and propagation and 
the transmission of outflow energy and momentum to the ambient medium.
However, the magnetic field structures have not always been observationally clear in/around cluster forming regions, 
particularly around nearby cluster forming regions because of the lack of deep, wide-field near-infrared (NIR) 
polarimetry data. 
 
The Serpens cloud core is one of the nearby\footnote{We assume a distance of 
$\sim$260 pc for the Serpens cloud, following the most of the recent papers on the Serpens cloud
and based on the discussion of \cite{st03} on the center distance of the Aquila Rift system.}, 
active  low-mass star forming regions  
at the northern part of the Serpens cloud and many observational works have been 
done \citep[][and references therein]{ei08}.  
Recent mid-IR studies \citep[e.g.,][]{ka04, ha07, wi07} revealed that a lot of embedded 
young stellar objects (YSOs), 
including Class 0/I objects, are located toward an aggregate of (sub)millimeter dust continuum cores 
 \citep[e.g.,][]{da99, ka04, en07}, which consists of two sub-clumps \citep[NW and SE sub-clumps;][]{ol02}
in the central region and is enveloped by ambient molecular gas  
\citep[e.g. $^{13}$CO, and C$^{18}$O;][]{mc00, ol02}.

Many outflow activities that are related to star formation have been  taking place
in the Serpens cloud core. 
CO high velocity flows are reported to be widely spread over 
the cloud core \citep[e.g., ][]{wh95, da99, nar02}. 
Compact molecular outflows of higher density tracers and H$_2$ jet-like knots 
are associated with the submillimeter cores 
\citep[e.g.,][]{cu96, her97, wo98, ho99, wil00}.   
The direction of these compact outflows was reported to be PA$\sim155\degr$ on an average 
with deviation of  a few $10\degr$ \citep[see Table 5 of][]{ol02}, 
which is nearly parallel to the alignment direction, from NW to SE, of the two sub-clumps
\citep[e.g.,][]{da99, ka04, en07},  
or to the cloud elongation in $^{13}$CO, C$^{18}$O, and other higher density tracers 
\citep[e.g. ][]{mc00, ol02}.
In addition, \cite{da99} and \cite{zi99} found, through optical narrow-band imaging,  
that many HH objects emanate from the two sub-clumps 
to the ambient region, penetrating the dense part of the central region. 

The Serpens Reflection Nebula (SRN) illuminated by SVS 2 \citep{st76} has been 
extensively studied by polarimetric measurements both in  optical and 
near-infrared (NIR)  wavelengths  \citep{ki83, wa87, go88, so97, hu97}. 
NIR polarimetric measurements \citep{so97, hu97} probed also some other obscured 
reflection nebulae around SVS 2 in detail.
\cite{go88} suggested the magnetic field of a NW-SE direction 
based on the elongation of the reflection nebulae around several YSOs in the central region of Serpens cloud core. 
In contrast, the NIR polarization measurement of a background star candidate suggested rather different 
direction of magnetic field because its polarization angle was nearly perpendicular to the NW-SE direction  \citep{so97}. 
However, this measurement was only for one background candidate, which in fact has a possibility of being a YSO 
in the Serpens cloud core and its polarization originating from the YSO itself. 
Therefore, it is vitally important to measure more background stars to resolve this discrepancy 
and to know the magnetic field structure toward the Serpens cloud core.

We conducted deep, $JHK$s imaging polarimetry of the Serpens cloud core to reveal 
the magnetic field structure in this region.   
We also searched for more NIR reflection nebulae associated with YSOs.
Here, we present the results of our imaging polarimetry in the Serpens cloud core 
by comparing the data from the previous observations and 
discuss the role of the magnetic field in this region. 

\section{Observations and Data Reductions}

\begin{figure}
\plotone{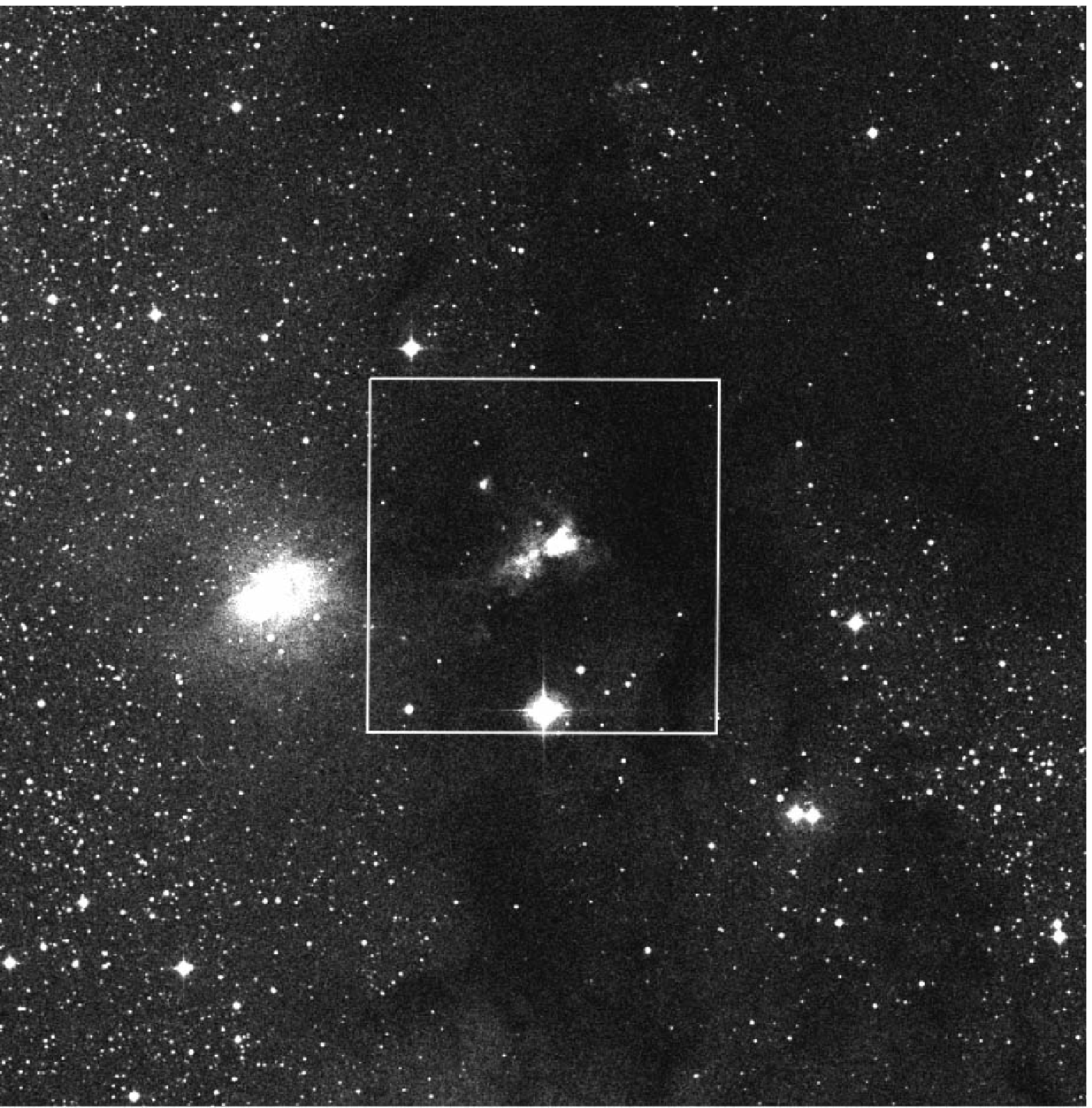}
\caption{DSS II R image of the Serpens cloud core.  The area 
of the polarimetric imaging is indicated  
by a rectangle with a size  of $8.4\arcmin \times 8.5 \arcmin$\label{f1} 
and  a center position of $(\alpha, {\mathit \delta})_ {\rm J2000}$ 
=(18:29:57.6, +01:14:34).}
\end{figure}

Toward the Serpens cloud core (Figure \ref{f1}), simultaneous $JHK$s polarimetric 
observations  were carried out on 2006 June 23 UT with the imaging polarimeter 
SIRPOL \citep{kan06}, which is an attachment of the near-infrared camera SIRIUS  
mounted on the IRSF 1.4-m telescope at the South Africa Astronomical Observatory.  
The SIRIUS camera is equipped with three 1024 $\times$ 1024 HgCdTe (HAWAII) arrays, 
$JHK$s filters, and dichroic mirrors, which enables simultaneous $JHK$s 
observations \citep{na99, na03}. 
The field of view at each band is $\sim$7\farcm7 x 7\farcm7 with a pixel scale of 
0\farcs45 pixel$^{-1}$.  

We obtained 10 dithered exposures, each 10 s long, at four wave-plate angles 
(0$\degr$, 22.5$\degr$, 45$\degr$, and 67.5$\degr$ in the instrumental 
coordinate system) as one set of observations and repeated this 9 times.  
Sky images were also obtained in between target observations.
Thus, the total on-target exposure time was 900 s per wave-plate angle. 
The seeing was $\sim1.2\arcsec$ at $K$s during the observations.  
Twilight flat-field images  were obtained at the beginning and end 
of the observations.

Standard procedures, dark subtraction, flat-fielding with twilight-flats, bad-pixel substitution, 
sky subtraction, and averaging of dithered images, were applied with IRAF. 
We first calculated the Stokes parameters  as follows; 
 $Q=I_{0} - I_{45}$, $U=I_{22.5} - I_{67.5}$, $I=(I_{0} + I_{22.5} + I_{45} + I_{67.5})/2$, 
 where $I_{0}$, $I_{22.5}$, $I_{45}$, and  $I_{67.5}$ are intensities at four wave-plate angles. 
To obtain the Stokes parameters in the equatorial coordinate system,  
a rotation of 105$\degr$  \citep{kan06}  was applied to them.  
We calculated the degree of polarization $P$, and the polarization angle $\theta$ as follows; 
$P=\sqrt{Q^2+U^2}/I$, $\theta=(1/2)\tan^{-1} (U/Q)$.
The polarization intensity ($PI$) is obtained by multiplying the total intensity ($I$) 
by the degree of polarization ($P$).
The absolute accuracy of the position angle of polarization was estimated to be 
better than 3$\degr$ at the first light observation of SIRPOL \citep{kan06}.
The polarization efficiencies are
95.5\%, 96.3\%, and 98.5\% at $J$, $H$, and $K$s, respectively, and 
the instrumental polarization is less than 0.3\% all over the field of view 
at each band \citep{kan06}.
Due to these high polarization efficiencies and low instrument polarization,  
no particular corrections were made here.

Aperture polarimetry was performed for $H$ and $K$s band point sources detected 
by DAOFIND in the field of view.  
No polarimetry for J band sources was done due to their much smaller number, 
compared with those of $H$ and $K$s band sources (see Figure \ref{f2} a, c, and e). 
APHOT of the DAOPHOT package was used to evaluate the point source magnitudes 
for four wave-plate angles at  $H$ and $K$s.  
An aperture radius of 3 pixels was adopted for each band.  
The errors of the degree of polarization (${\mathit \Delta} P$) and the position angle were 
calculated from the photometric errors, and the degrees of polarization were debiased 
as $P_{\rm debias}=\sqrt{P^2 - {\mathit \Delta} P^2}$ \citep{wa74}. 
Hereafter, we use $P$ as substitute for this debiased value for the aperture photometry data.
Only the sources with photometric errors of $<$0.1 mag and 
$P/{\mathit \Delta} P > 2$ were used for analysis. 
The 2MASS catalog \citep{sk06} was used for absolute photometric calibration. 
The limiting magnitudes at 0.1 mag error level
were estimated to be 18.6 at  $H$  and 17.5 at $K$s.   

\section{Results}
\subsection{Polarizations of extended emission}

\begin{figure}
\begin{center}
\includegraphics[scale=.40]{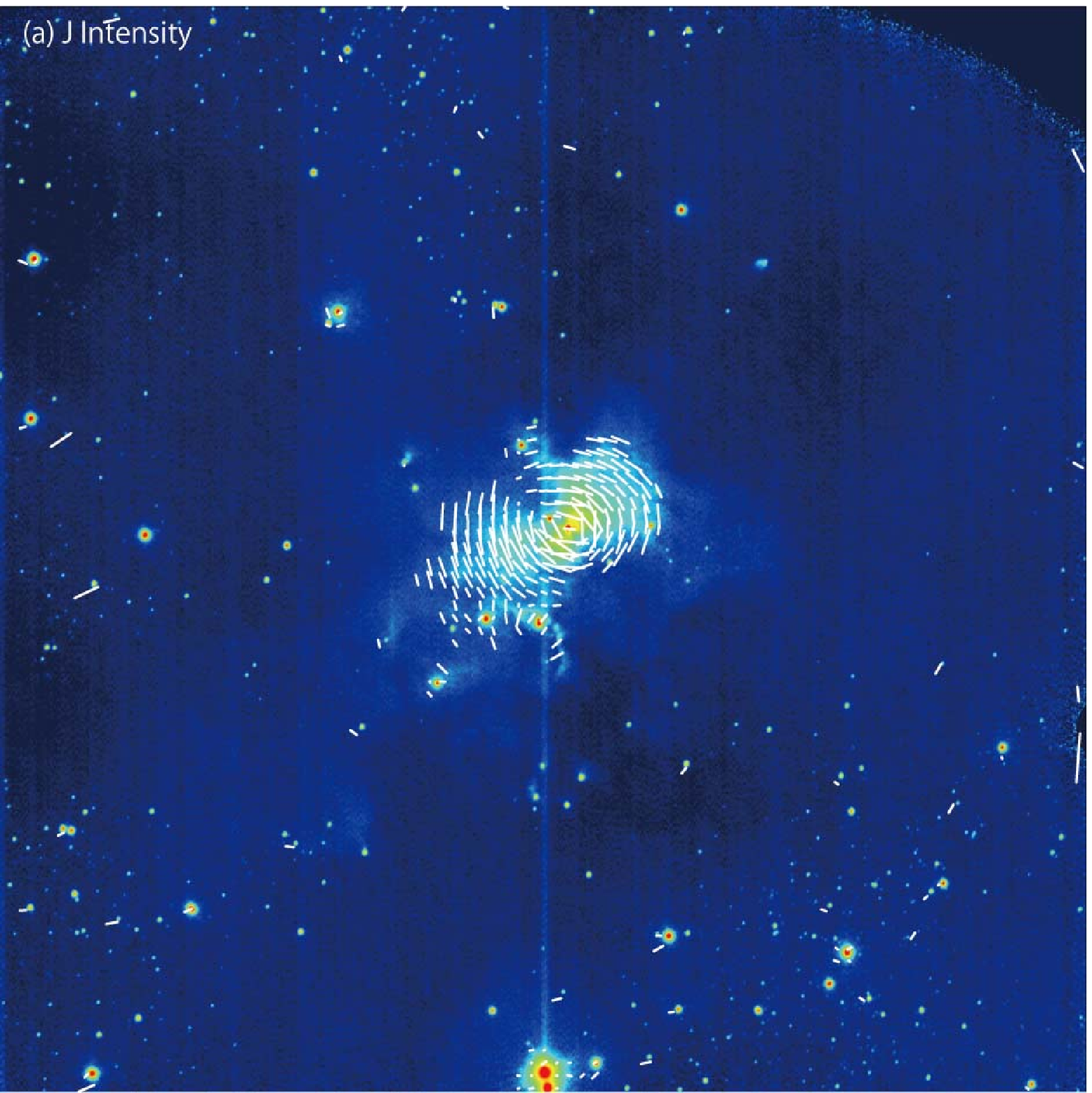}
\includegraphics[scale=.40]{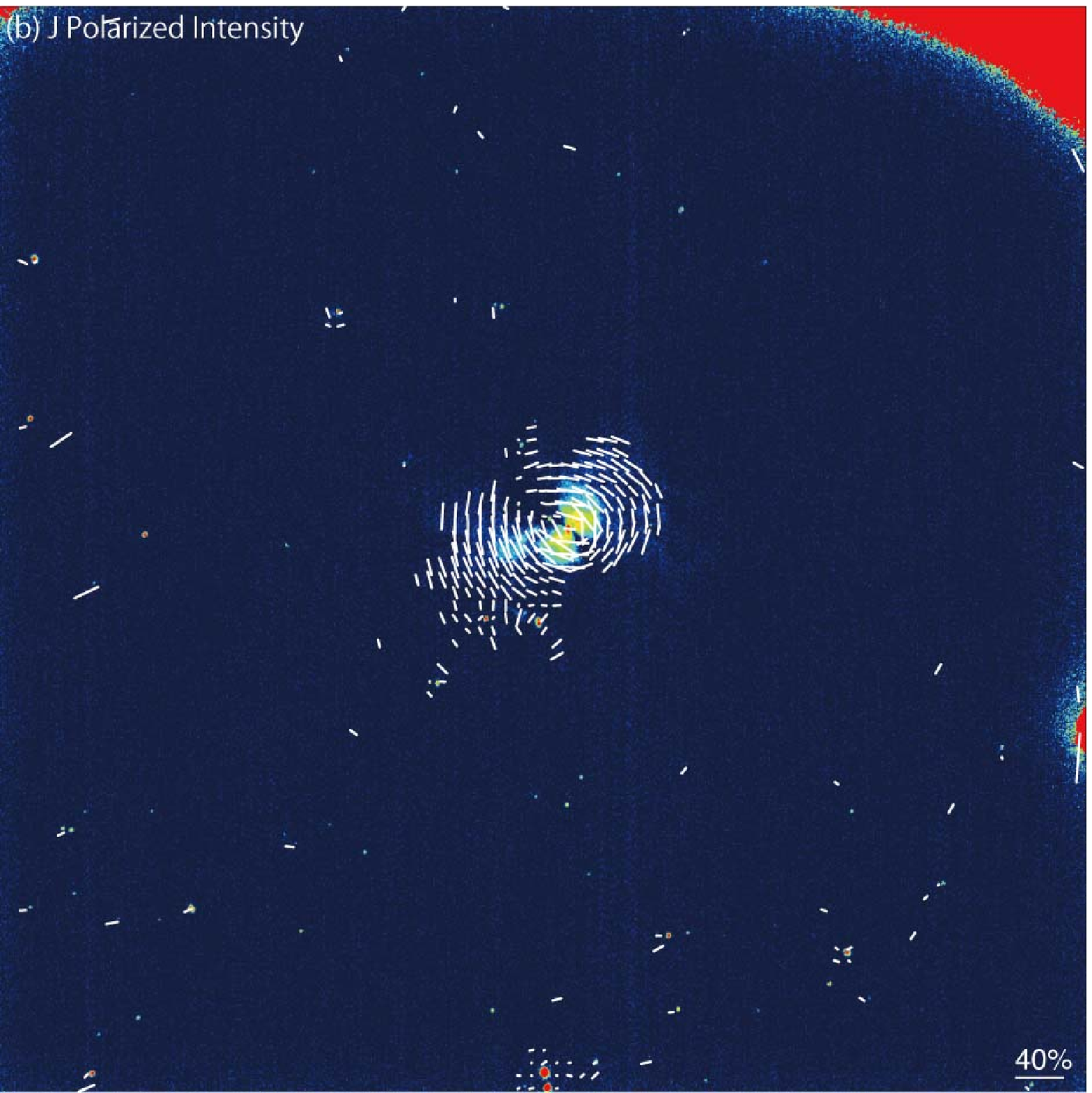}
\includegraphics[scale=.40]{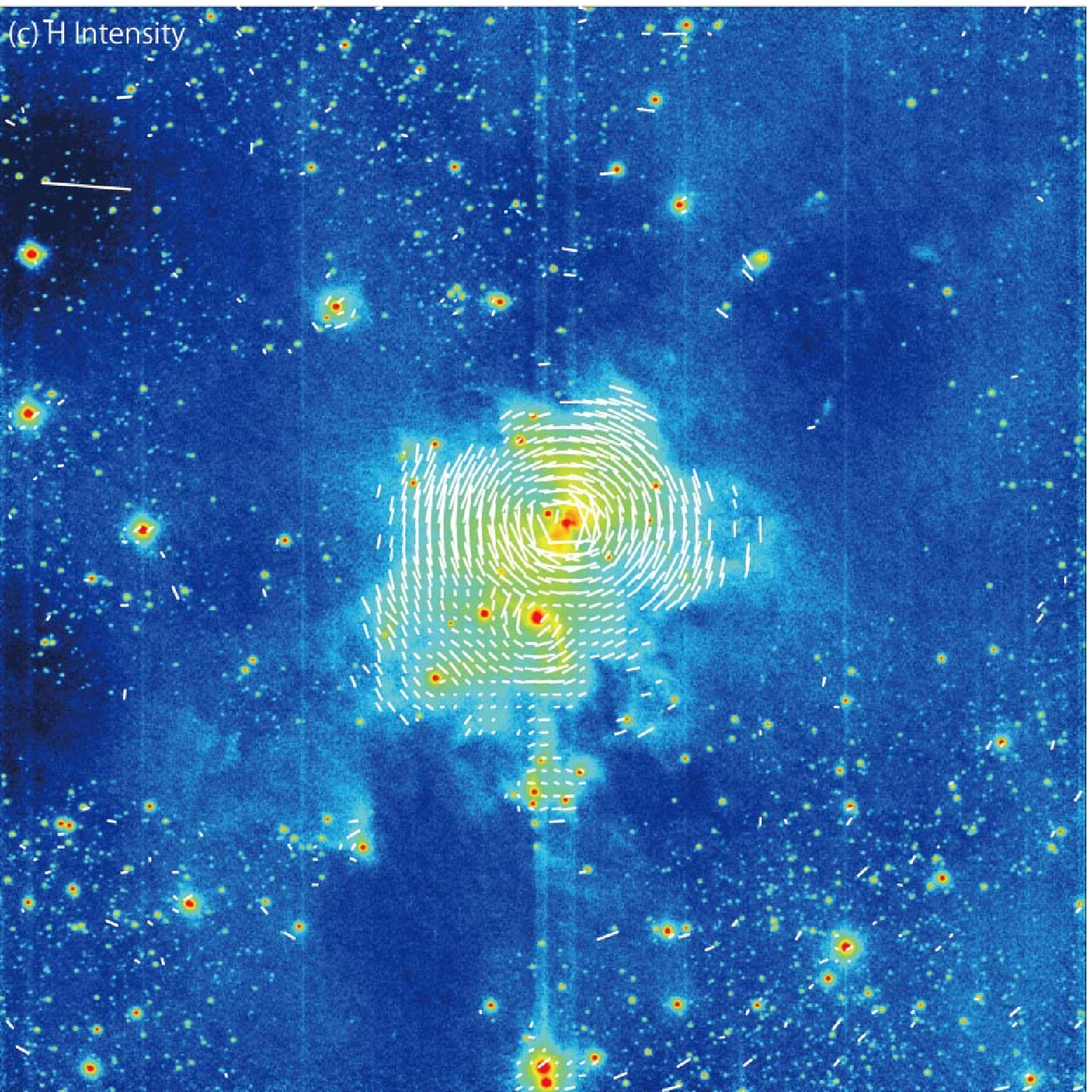}
\includegraphics[scale=.40]{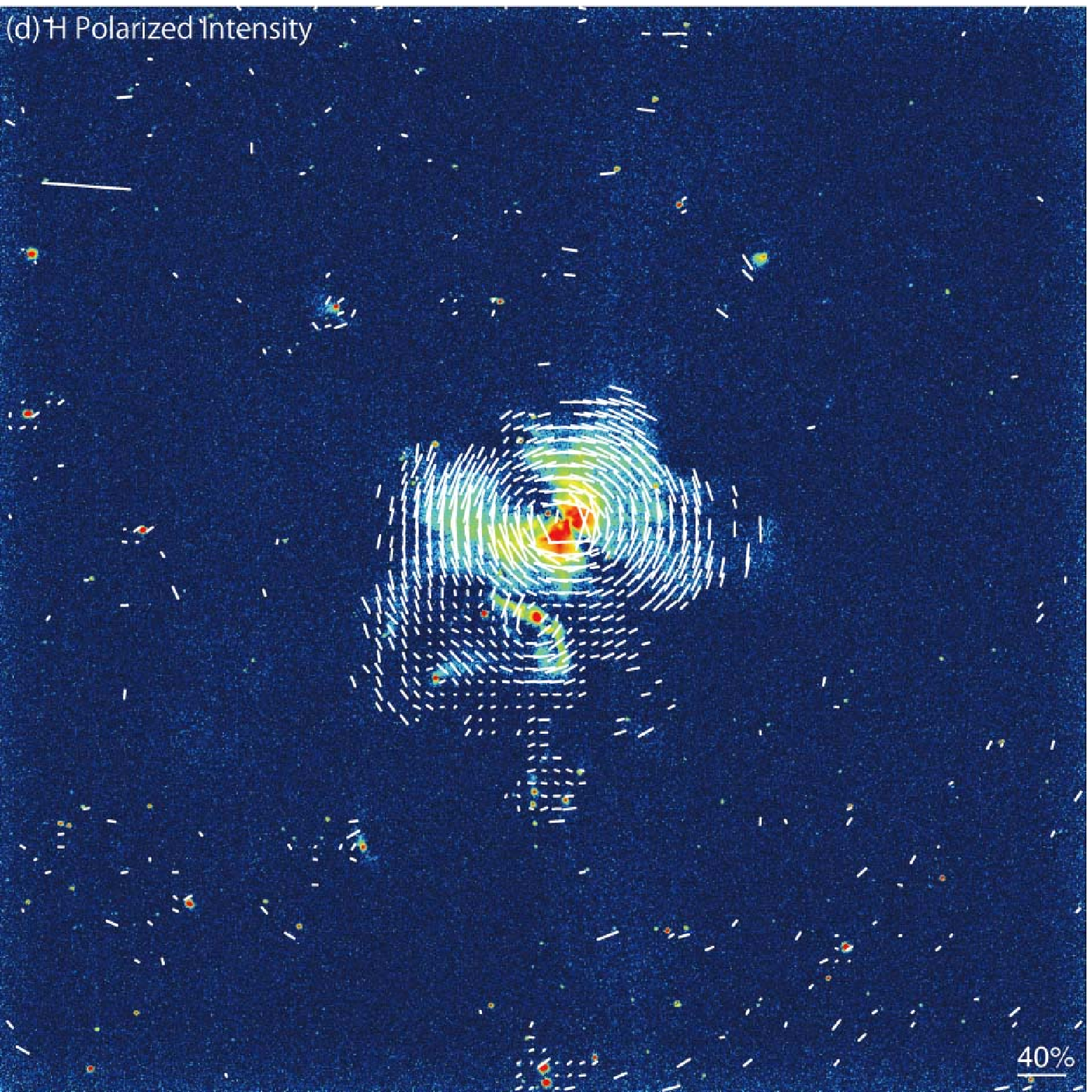}
\includegraphics[scale=.40]{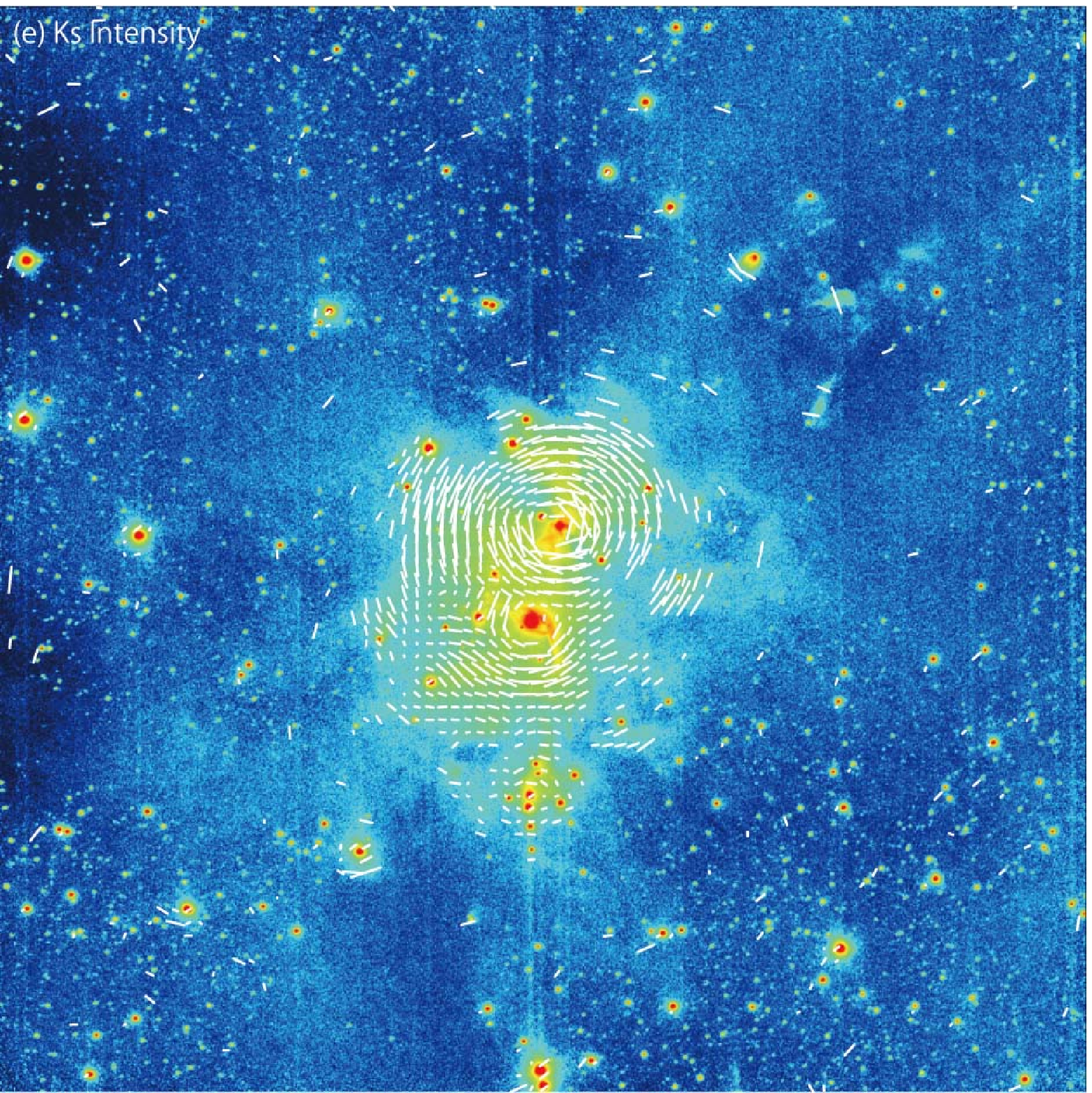}
\includegraphics[scale=.40]{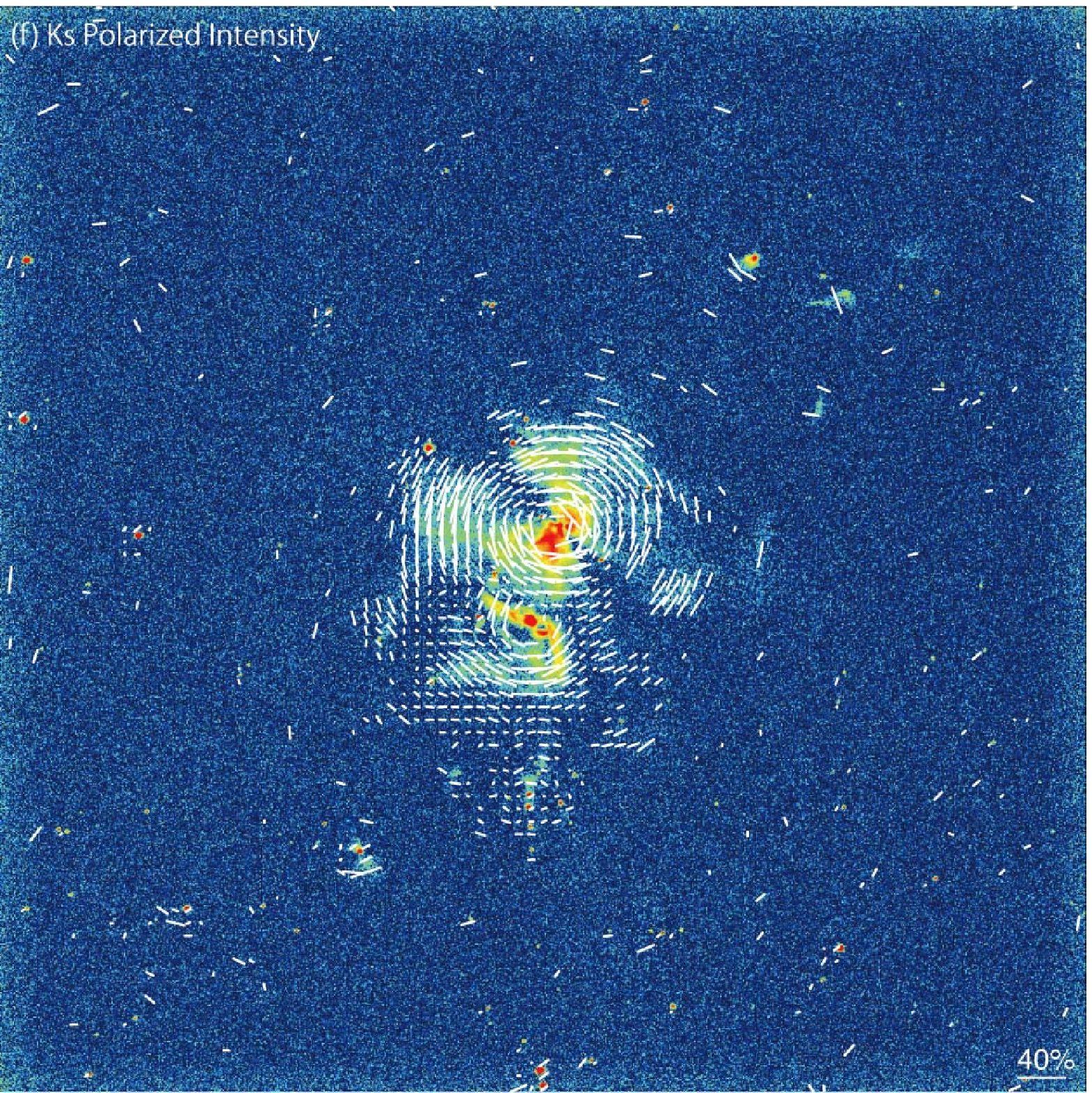}
\end{center}
\caption{$JHK$s polarization vector maps of the Serpens cloud core, 
superposed on the total and polarized intensity images in the logarithmic scale. 
The vectors were made by 12$\times$12 pixel binning. 
The field of view is $\sim7\farcm7 \times 7 \farcm7$.  
North is at the top and east is to the left.  
The center position of each image is $(\alpha, {\mathit \delta})_ {\rm J2000}$ 
=(18:29:57.39, +01:14:36.2). 
The dead pixel regions of the $J$-band images are at the upper right and 
near the middle of the right edge.\label{f2}}
\end{figure}

The $JHK$s polarization vector maps of the Serpens cloud core 
are shown, superposed on the total and polarization intensity images  in Figure \ref{f2}.  
These polarization maps clearly indicate that the central part of the reflection nebula 
is illuminated mainly by two sources; the north part (SRN) by SVS 2  
and the south part by SVS 20,  at $H$ and $K$s with two centrosymmetric patterns 
\citep[see also][]{so97, hu97}, 
while at $J$ only SRN is dominant  \citep[see also][]{so97} 
as is seen in the optical \citep{wa87,go88}. 
This invisibleness at shorter wavelengths suggests 
that the southern part of SRN around SVS 20 
is more obscured than the northern part around SVS 2, 
consistent with the $A_V$ map deduced from $H-K$ color \citep{hu97}.
The centrosymmetric patterns are more clearly shown in Figure \ref{f3}, 
which is a $K$s polarization vector map shown with a resolution higher than Figure \ref{f2}e. 

The $PI$  images and vector patterns of SVS 2 clearly show that 
SVS 2 is associated with a bipolar structure with a dark lane.
In the $JHK$s intensity images of Figures \ref{f2} a, c, and e,  
at shorter wavelengths, the NW lobe of the bipolar nebula is brighter 
than the SE lobe, while at longer wavelengths the SE lobe is brighter.  
This suggests that the NW lobe  is near side and that the SE lobe is far side.
The nebula structure and  dark lane of SRN have been already reported 
in the two polarimetric studies \citep{so97, hu97}.  
In addition,  \cite{po05} modeled SRN as a disk shadow system with their imaging data,
and suggested that SVS 2 is associated with a small disk, which is not unresolved, 
and a spherically symmetric envelope.  

The nebula illuminated by SVS 20 is clearly recognized at $H$ and $K$s 
with a centrosymmetric pattern around SVS 20.  
This object has a peculiar morphology with a ring and 
two arms protruded from that ring. 
Because we plan to report its details in a separate paper, 
including other YSOs with reflection nebulae,  we will not mention the details here. 

\subsection{Polarizations of nebulosities associated with YSOs}
\subsubsection{The central region}

Figure \ref{f3} presents an higher resolution $K$s polarization 
vector map with 3$\times$3 pixel binning toward the central region 
of the overall image, superposed on the $K$s intensity map.  
With this map and/or the highest resolution vector maps without binning, 
we identified stellar sources having reflection nebulae locally illuminated by themselves  
with centrosymmetric or centrosymmetric-like patterns.
It is not easy to identify such sources 
only from Figure \ref{f3} due to the strong contamination from SVS 2 and SVS 20. 
It is also not easy toward the SVS 4 cluster, which is a compact cluster located 
to the south of SVS 20, due to the source congestion. 
We therefore used the highest resolution vector maps without binning
for the sources having the strong contamination (see Appendix).
The identified sources with reflection nebulae illuminated by themselves are 
marked in Figure \ref{f3}, including SVS 2 and SVS 20, and are listed 
in Table \ref{tbl_a}.  

Most of the identified sources are relatively bright in the central region. 
This is probably due to the strong light from SVS 2 and SVS 20 and only brighter 
sources with reflection nebulae may be detected.  
Except EC117 (SSTc2dJ 18300065+0113402), all the identified sources are classified 
as sources that have outer disks with an excess at least  8 and/or 24 $\micron$, 
i.e., class 0/I, flat spectrum, class II and transition disk sources \citep{wi07}.
Although EC117 is classified as a class III source without an outer disk  due to 
no detection of $24 \micron$ continuum \citep[Table 4 of ][]{wi07},   
it was reported that EC117 has 
a flux of 3.20$\pm$1.43 mJy at $24 \micron$ \citep[Table 3 of][]{ha07}.
This could suggest the outer disk of EC117, but the signal to noise ratio of 
$\sim$2 is not high enough for the robust detection at $24 \micron$.  
Almost all members of the SVS 4 cluster seem to be associated with reflection nebulae.

\subsubsection{The NW region}

Figure \ref{f4} presents a high resolution $K$s polarization 
vector map without binning toward the NW region of SRN.  
Here we identified stellar sources associated with self-luminous nebulae
as well as those with reflection nebulae by using this map and listed them in Table \ref{tbl_b}.
Three sources, DEOS, EC53, and EC67, are associated with reflection nebulae 
having centrosymmetric or partially centrosymmetric vector patterns.  
The other sources are associated with elongated nebulae or jet-like knots 
emanating from the sources in a straight line and their polarization 
vectors are almost perpendicular to their elongation directions.   
The elongated structures or knots are likely to be created/excited 
by outflows from these sources.  

The jet-like knots are clearly seen near the north-west of EC41, 
which was considered to be an embedded star 
but not a driving source of this jet \citep{ei89, ho99}.  
These jet-like knots are reported to be mostly H$_2$ emission with  
weak continuum \citep{ho99}, and their polarization of these knots is 
nearly perpendicular to the jet elongation, though the polarization directions are more 
scattered in the northern knots than in the southern knots of this jet-like structure.  
At $H$-band, the polarization vectors of weak continuum emission of the southern knots 
are also nearly perpendicular to the jet elongation, 
which is parallel to the radial direction from SMM1-FIRS1, not from EC41.
Thus, SMM1-FIRS1 is the illuminating source of these jet-like knots, 
and the jet-like knots could correspond to the cavity walls that were created 
by the outflow from SMM1-FIRS1.  

The jet-like structure from SMM1-FIRS1 seems to continue farther away to 
a bow-shock-like nebulosity located at $\sim$80-90$\arcsec$ north-west of SMM1-FIRS1 
(or at $\sim20\arcsec$ north-west of EC28). 
The polarization vectors at this bow nebulosity indicate either SMM1-FIRS1 or EC28 
is illuminating or exciting it.
No information is available on whether the bow nebulosity is really shock-excited 
H$_2$ emission or not, due to its position outside Figure 3 of \cite{ho99}, 
although week $H$-band continuum emission is detectable with polarization angles 
similar to those of $K$s-band.
The alignment with the jet structure, the bow nebulosity and HH 460, 
which is located at $\sim4 \arcmin$ north-west to SMM1-FIRS1 \citep{da99}, 
gives a hint that the bow nebulosity is related to the outflow from SMM1-FIRS1.  
The associations of the blueshifted CO lobe with HH 460 \citep{da99} and 
of the bow nebulosity with the CS emission \citep[CS1;][]{te00}, which is considered to be related 
to the outflow, support the shock excitation of the bow nebulosity.  
However, it is impossible to completely exclude the possibility that EC28, 
which is the closest NIR source to the  bow nebulosity, or SMM1-FIRS1 itself
contributes to the illumination of the bow nebulosity,  
due to the scattering of the polarization vector directions. 
In the midway from SMM1-FIRS1 to this bow nebulosity, there exist some faint knots 
that are almost H$_2$ emission \citep[see Figure 3 of][]{ho99}, 
but no polarization is detectable for these knots. 

The chain of nebulosity knots, located just south-east to the 3 mm continuum core S68Nc 
\citep{te98, wil00}, was reported to be H$_2$ emission jets that originate from 
the 3mm subknot a3/S68Nc \citep{ho99}.  
Although only several polarization vectors are shown toward these knots, 
they could imply that their polarization direction is nearly perpendicular to the jet elongation.

A nebulosity protruding from EC38/S68Nb is seen, and its polarization vectors appear 
to be nearly perpendicular to the protruding direction.
A faint, small, elongated nebulosity is recognized just south-east to SMM10-IR. 
Although some nebulosities illuminated from SVS 2 are also seen near SMM10-IR, 
this nebulosity is most likely a nebulosity related to SMM10-IR based on its morphology. 
No information is available on whether these two nebulosities are shock-excited 
H$_2$ emission or not, because they are out of Figure 3 of \cite{ho99}. 
We note that no $H$-band emission is detectable toward SMM10-IR, 
while very week $H$-band continuum is seen toward the nebulosity protruding from EC38/S68Nb. 		

\subsection{Polarizations of point sources}

We have measured $H$ and $K$s polarization for point sources, 
in order to examine the magnetic field structures.
Only the sources with photometric errors of $<$0.1 mag and 
$P/{\mathit \Delta} P > 3$ were used for analysis. 

The top panel of Figure \ref{f5} presents the polarization degrees at $H$ versus 
$H-K$s color diagram for sources having polarization errors of $<0.3\%$.  
YSOs identified by \cite{wi07} are not included in this diagram.  
In this diagram, the maximum of polarization degree at an $H-K$ color is roughly 
proportional to the $H-K$ color, i.e., the extinction, having consistency that the 
origin of the polarization is dichroic absorption.  
Therefore, we consider the polarization of these point sources as the polarization 
of the dichroic origin, and that their polarization vectors represent the directions 
of the local magnetic field averaged over their line of sight of the sources. 
In the nearby star forming regions such as the Taurus and Ophiuchus clouds, 
the highest value of the maximum polarization efficiency was reported 
to be $P(H)/E(H-K)=4.6$ or $P(H)/A(H)=1.6$ \citep{ku08}, 
which were derived from the data of \cite{wh08}. 
In Figure \ref{f5}, a dashed line represents $P(H)/(H-K\rm{s})=4.6$ 
where the offset of the intrinsic $H-K\rm{s}$  color is ignored.  
Our sources have the maximum polarization efficiency of $P(H)/(H-K\rm{s})=6.2$ 
(thick line) similar to that of the  nearby star forming regions, and this is also 
consistent with the dichroic origin. 

The bottom panel of Figure \ref{f5} shows the $H$-band polarization angles 
of the point sources with $P <  6.2(H-K\rm{s})$, of which the polarization vectors are shown 
in Figure \ref{f6}. YSOs are not included in the bottom panel, but included in Figure \ref{f6}. 

The polarization angles are mostly in a range of $\sim$0--140$\degr$ 
and their median and average angles are  63.5$\degr$ and 64.6$\pm$35.6$\degr$, respectively. 
While the polarization angles are largely scattered, there is a tendency that the degree of scatter becomes smaller in the redder  $H-K\rm{s}$  color region.
This tendency suggests that the polarization angles are more confined 
in the inner region (redder color region) than in the outer region.  

The magnetic field is neither simply straight nor random over the whole field (Figure \ref{f6}). 
The vectors appear to be systematically ordered and gradually curved, 
suggesting a clear hourglass shape that is left-handedly tilted by $\sim 60$--$80\degr$ 
and the direction of the global magnetic field that is nearly perpendicular to the elongation of 
the Serpens cloud core from NW to SE, $\sim150\degr$ 
\cite[e.g., the C$^{18}$O maps of][]{mc00, ol02}. 

Signs of hourglass shapes in  the magnetic field have been already reported 
in high-mass star forming cores such as NGC 2024 \citep{la02},  
OMC-1 \citep{sc98, hou04a, ku08}, and DR21 Main \citep{ki09}.  
In low mass cores such as NGC 1333 IRAS 4A \citep{gi06} and  Barnard 68
\citep{kan09} hourglass shapes have been more clearly shown.  
These examples, except OMC-1, of the hourglass-shaped magnetic field have been found only 
in isolated cores or cores with simple structures in the star forming regions. 
However the Serpens cloud core is a molecular cloud complex consisting of many molecular 
cloud cores or sub-millimeter cores \cite[e.g.,][]{da99}, which form a cluster of low-mass YSOs,  
and  the hourglass-shaped magnetic field spreads widely over the Serpens cloud core.
Thus, this is a clear example that the hourglass-shaped magnetic field is 
associated with a low-mass star forming complex,
while OMC-1 is an example of a hight-mass star forming complex.

\section{Analysis and Discussion}

\subsection{Shape of the magnetic field}

We have modeled the shape of the magnetic field with the polarization vectors measured 
at $H$ for point sources, following \cite{gi06} and \cite{kan09}.   
The magnetic field was fitted with a parabolic function of $x=g+gCy^2$, 
with a counterclockwise tilted $y$-axis (the parabolic magnetic field axis of symmetry) 
by $\theta_{\rm PA}$ and a symmetric center $(x, y)_{\rm center}$,  
where the $y$ is the distance from the horizontal axis ($x=0$) 
and the $x$ is the distance from the parabolic magnetic field axis of symmetry. 
The value of $\tan^{-1}(dy/dx)+90\degr$ corresponds  
to the position angle of the polarization ($\theta$).   
Only the point sources, except YSOs, having $P/{\mathit \Delta}P > 3$ and $P < 6.2 (H-Ks)$, 
were used for the fitting.
The error of the polarization angle (${\mathit \Delta}\theta$) 
was used to compute a weight for the datum, $1/({\mathit \Delta}\theta)^2$.

In Figure \ref{f7}, the best-fit magnetic field is shown 
as well as the measured polarization vectors for 149 sources. 
The position angle of the parabolic magnetic field axis of symmetry
is $\sim70\degr$, and the coefficient $C$ of $y^2$ is $\sim7.1\times 10^{-6}$ pixel$^{-2}$.
The root mean square  (r.m.s.) of the residuals is $\sim22\degr$.

We executed one-parameter fitting of the magnetic field in local areas,  
in order to more accurately calculate the r.m.s.  of the residuals, 
with the same $\theta_{\rm PA}$ and $(x, y)_{\rm center}$ 
obtained in the global fitting above. 
We selected three corners and one more area of the image 
where the source density is relatively high and/or the magnetic field seems to 
be rather ordered (areas outlined by dashed boxes in Figure \ref{f7}).    
Toward the SE corner of the image ( $x<400$ and $y<400$ in Figure \ref{f7}; 30 sources), 
the coefficient $C$ of $y^2$ was determined to be 
$(7.99\pm 0.76)\times10^{-6}$ pixel$^{-2}$, similar to that of the global fitting,  
and the r.m.s. of the residual was calculated to be $12.9\pm 0.9\degr$, 
and toward the SW corner ($x>500$ and $y<230$; 20 sources),  
$C=(7.52\pm 1.00)\times 10^{-6}$ pixel$^{-2}$ and r.m.s. = $27.0\pm2.0\degr$ 
were obtained.  
Removing the dispersion due to the measurement uncertainties of the polarization 
angles $4.2\pm3.0\degr$ and $3.2\pm 2.2\degr$, 
we obtained the dispersions from the best-fit model, 
$12.2\pm1.4\degr$and $26.8\pm2.0\degr$ for the SE and SW corners, respectively.
Toward the NE corner  ($x<300$ and $y>800$; 18 sources), 
$C=(3.36\pm 0.92)\times 10^{-6}$ pixel$^{-2}$ and r.m.s. = $14.8\pm 1.6\degr$ 
were evaluated, and the intrinsic dispersion from our model of $13.7\pm2.0\degr$  
was obtained with the measurement uncertainty of $5.6\pm2.4\degr$. 
This smaller $C$ indicates that the curvature of the magnetic field here is rather looser 
than that expected from the global fitting, i.e., slightly bended to the direction parallel to 
the symmetry axis of the magnetic field. 
Toward the area next to the NE corner ($400<x<700$ and $y>800$; 17 sources), 
$C=(6.85\pm 0.55)\times 10^{-6}$ pixel$^{-2}$ and r.m.s. = $13.3\pm1.4\degr$ 
were evaluated, and the intrinsic dispersion of $12.6\pm 1.8\degr$ was obtained with the measurement uncertainty of $4.2\pm 2.9\degr$. 

\subsection{Comparison of the magnetic field with the submillimeter and millimeter data}

\subsubsection{850 $\micron$ continuum}

We compare our $H$-band measured polarization vectors and the modeled magnetic field 
with the 850 $\micron$ dust continuum map of \cite{da99} in Figure \ref{f8}.
Note that the green lines of this figure do not present lines of magnetic force, 
just the direction of the magnetic field.
   
The high intensity ridge of  the 850 $\micron$ continuum is elongated along the NW-SE direction, 
having two sub-clumps (NW and SE sub-clumps), 
both of which consist of several dense cores (e.g., SMM 1--11, S68Nb--d, and PS2 in Figure \ref{f8}).  
This distribution of the 850 $\micron$ continuum is very similar to that of the bright parts 
of the $^{13}$CO($J=1-0$) and C$^{18}$O($J=1-0$) emission \citep{mc00, ol02}, 
although the global distribution of the $^{13}$CO($J=1-0$) emission is not always elongated, 
but rather roundly extended \citep{ol02}.
It is evident that the symmetric axis ($y'$-axis) of the best-fit magnetic field 
with a parabolic function is nearly perpendicular to the elongation direction of  
these continuum and molecular line emissions.  
The horizontal axis ($x'$-axis) of the parabolic magnetic field
is situated nearly along the  850 $\micron$ continuum ridge, 
although there are some deviations of the continuum emission from the horizontal axis.  
The symmetric axis of the parabolic magnetic field runs through the northern part of 
the SE sub-clump, not through the middle point of the two sub-clumps, 
which looks like the center of gravity of the Serpens cloud core 
when we glance at the 850 $\micron$ continuum map.  

\cite{da99} suggested the presence of three extended cavity-like structures 
to the east of SMM 3 (hereafter CLS 1), south-west of SMM 2 (hereafter CLS 2), 
and north-west of SMM 4 (hereafter CLS 3), which consist of three pairs of 
filaments that protrude the 850 $\micron$ continuum ridge. 
They mentioned that these cavity structures (CLS 1--3) are probably shaped 
by outflows rather than by global cloud collapse along, say, magnetic field lines.

As is in Figure \ref{f8}, the filaments to the north-east of SMM 3 and east of SMM 2 form 
CLS 1, those to the south-east of SMM2/PS2 and south of SMM11 form CLS 2, 
and those to west of SMM3 and east of SMM4 form CLS 3.  
It appears that the two filaments of CLS 1 jut almost along the magnetic field 
from the SE sub-clump and that the symmetry axis ($y'$-axis) of the magnetic field 
go through the inside of CLS 1 as well as CLS 3.

\subsubsection{CO emission}

Here we compare our best-fit magnetic field with the $^{12}$CO $J=2-1$, 
$^{12}$CO $J=1-0$, $^{13}$CO $J=1-0$, and C$^{18}$O $J=1-0$ 
observations \citep{wh95, da99, mc00, nar02, ol02}.  

\subsubsubsection{$^{12}$CO $J=2-1$}

As was mentioned above, the bright parts of the $^{13}$CO $J=1-0$ and 
C$^{18}$O $J=1-0$ emission maps are elongated and confined in the ridge, 
while the global distributions of $^{12}$CO $J=2-1$ and $^{13}$CO $J=1-0$, i.e., 
the low density molecular gas, are extended \citep[e.g.,][]{wh95,da99,ol02}. 

Figure \ref{f9} presents our best-fit magnetic field superposed on the CO $J=2-1$ 
contour map and 850 $\micron$ image of \cite{da99}, 
where the CO $J=2-1$ map is considered to show the ambient molecular gas 
of the Serpens cloud core, but not the dense cores. 
\cite{da99} mentioned that toward the two filaments of CLS 1 and 
one CLS 2 filament to the south-east of SMM2/PS2 
the CO $J=2-1$ emission and 850 $\micron$ continuum distributions coincide well. 
As mentioned above, the two filaments seem to run almost along the magnetic field, 
indicating that the CO $J=2-1$ filaments are also related with the magnetic field. 
For the CLS 2 filament to the south-east of SMM2/PS2, 
the same situation as the CLS 1 filaments may be also seen. 
Two other CO $J=2-1$ filaments/extensions to the north-west of SMM 9 and west of 
SMM 1 are also noticeable in Figure \ref{f9}. 
Although considerable parts of these two filaments are out of our polarimetry image, 
the extrapolation of our best-fit magnetic field cloud predict that these two filaments run 
along the magnetic lines. 

\subsubsubsection{C$^{18}$O $J=1-0$ and $^{13}$CO $J=1-0$}

\cite{mc00} showed that a velocity gradient running from a LSR velocity centroid of 9 km s$^{-1}$ 
at the north-west end of the C$^{18}$O $J=1-0$ emission to 7.5 km s$^{-1}$ at the south-east end 
\citep[Figure 2 of][]{mc00}, i.e., along the elongation direction of C$^{18}$O.  
On the other hand, \cite{ol02} suggested that the Serpens cloud exhibits a velocity gradient 
roughly from east to west, based on their model fitting of velocity gradients in C$^{18}$O $J=1-0$, 
$^{13}$CO $J=1-0$, C$^{34}$S $J=1-0$, adopting their map center, 
which is the middle point of the two sub-clumps, as the reference position for analysis.
However, according to their channel and centroid velocity maps \citep[Figures 7 and 8 of][]{ol02}, 
the bright parts of C$^{18}$O $J=1-0$ and $^{13}$CO $J=1-0$ 
are similar to that of \cite{mc00}, and a steep velocity gradient from NW to SE 
almost along the normal line of the symmetry axis ($y'$-axis) of the magnetic field can be seen 
at just south of their reference position in $^{13}$CO, 
although at the reference position a velocity gradient from West to East is seen. 
It is surprising that the normal line of the steep velocity gradient almost coincides 
with the symmetry axis ($y'$-axis) of the magnetic field.  

In summary, the direction of velocity gradient is nearly along the elongation of 
the Serpens cloud core and is nearly perpendicular to the symmetry axis of the magnetic field 
with a coincidence of the normal line of the steep velocity gradient and the axis of the magnetic field. 
It could be possible that this normal line of the velocity gradient is an axis of the global rotation of 
the Serpens cloud core if the real center of gravity of the Serpens cloud core is located 
on the symmetry axis of the magnetic field.  

It is interesting to examine the presence of C$^{18}$O $J=1-0$ and $^{13}$CO $J=1-0$ features 
that coincide with the filaments of the CO $J=2-1$ emission and 850 $\micron$ emission.  
In the C$^{18}$O $1-0$ integrated emission maps of \cite{wh95}, \cite{mc00} and \cite{ol02}, 
a feature to the north-east of SMM 3 could coincide with one of the CSL 1 filament, 
but one to the east of SMM 2 is not clear.
In the channel map of $^{13}$CO $J=1-0$ \citep[Figure 7 of][]{ol02}, a filament feature 
to the east of SMM 2 is clearly visible in the blue-shifted emission at the panel of  
$V_{\rm LSR}$=5--7.3 km s$^{-1}$. 
This filament looks likely to coincide with the CSL 1 filament to the east of SMM 2, 
but we can clearly recognize that it is located just outside this CSL 1 filament, 
i.e., between this CSL 1 filament and the CSL 2 filament to the south-east of SMM 2/PS2.  
At the same panel, a feature to the north and north-east of SMM 3 or near SMM 8 is also visible. 
This feature appears to be just outside the CLS 1 filament to the north-east of SMM 3.  
At the panel of $V_{\rm LSR}$=8.4--12.4 km s$^{-1}$, a red-shifted feature that protrudes 
from the SE sub-clump is visible, but it is located toward the inside region of CLS 1.  
The presence of this red-shifted feature and the blue-shifted features are probably 
consistent with red-shifted velocity region that jut from the SE sub-clump 
and with blue-shifted regions toward both sides of this red-shifted region, 
respectively, in the $^{13}$CO $J=1-0$ centroid velocity map of \cite{ol02}.

\subsubsubsection{CO outflows}

\cite{da99} presented the integrated intensity contours of CO $J=2-1$ blue- and 
red-shifted outflows \citep[Figures 4 and 8 of][]{da99}. 
These figures imply that the 850 $\micron$ filaments that coincide the CO $J=2-1$ 
filaments are shaped by outflows. 
On the basis of a fact that these filaments run along the magnetic field, 
the outflows that protruded from the ridge to its ambient are most likely 
to be guided by the magnetic field or to drag the magnetic field. 
The outflows may be guided by the magnetic field since the magnetic field 
seems to be strong enough to be ordered at least over our polarimetric imaging area.   

The CLS 1 filaments are associated with red-shifted outflows, but no red-shifted 
CO $J=2-1$ emission is visible at the root of CLS 1.  
However, CO $J=1-0$ obervations \citep{nar02} showed U-shaped, 
red-shifted high velocity flow at the root of CLS 1. 
This CO $J=1-0$ feature and our best-fit magnetic field 
support the idea of \cite{da99} that the CLS 1 filaments of the CO $J=2-1$ and 
850 $\micron$ emission illustrate the action of a wide-angled wind powered by a source 
within the SMM 2/3/4 cluster, which has swept up gas and dust into a warm, 
compressed shell, although there is a possibility that the wind is powered by 
multiple sources within the cluster.

\subsection{Magnetic field strength}

We try to make an evaluation of the magnetic field strength toward four areas 
where we calculated the angular dispersions (residuals) for our best-fit magnetic field, 
using the Chandrasekhar \& Fermi (CF) method \citep{cha53}.  
On the basis  of the conclusions of recent MHD studies that the introduction of a correction 
factor is needed for evaluating the plane-of-sky component of the magnetic field 
\citep{os01,pa01,he01,ku03}, 
\cite{hou04b} mentioned that a correction factor of $\sim$0.5 is appropriate 
in most cases when the magnetic field is not too weak.  
Since the magnetic field seems to be ordered over the Serpens cloud core, 
the magnetic field is expected to be strong.
Therefore we first adopt a correction factor of 0.5 to evaluate the magnetic field strength. 
We need the mass density and velocity dispersion of the matter coupled to the magnetic field
to evaluate the magnetic field strength.  
Here, we use those estimated from the C$^{18}$O observation \citep{ol02}.  

Toward the four areas, the H$_2$ column densities from C$^{18}$O could be 
estimated to be $\sim 6\times10^{22}$ cm$^{-2}$ from Figure 11 of \cite{ol02}.  
Adopting the approximate C$^{18}$O extent of $\sim$12\arcmin  
\citep[$\sim$0.9 pc at d$\sim$260 pc; Figure 2 of ][]{ol02} as the depth of these area, 
we obtain the H$_2$ densities of $\sim 2.1 \times10^{4}$ cm$^{-3}$.  
From Figure 10 of \cite{ol02}, the C$^{18}$O velocity widths could be estimated to be 
$\sim$1.6--1.8 km s$^{-1}$ toward the SE and NE corners, and $\sim$1.8--2.0 km s$^{-1}$ 
toward the area next to the NE corner.   
Toward the SW corner with a complex distribution of velocity width, the velocity width may be 
$\sim$1--2 km s$^{-1}$.  
Using a mean molecular mass, $\mu$, of 2.3 and these values to derive the velocity dispersions, 
we roughly evaluated the magnetic field strength of the plane-of-the-sky of 
$B_{\parallel} \sim$160--180 $\mu$G toward the SE corner,  
$\sim$150--160 $\mu$G toward the NE corner, 
and $\sim$180--200 $\mu$G toward the area next to the NE corner. 
Although $\sim$50--90 $\mu$G can be evaluated toward the SW corner, 
this value might be more uncertain than those toward the other areas 
due to the larger uncertainty of the velocity width.  
The magnetic field strength evaluated here is higher than those measured 
around dark cloud complexes and prestellar cores, a few 10 $\mu$G  \citep[e.g.,][respectively]{alv08, kan09}, 
but smaller than those around HII regions, a few mG \citep[e.g.,][]{hou04b} 
and of a protostellar envelope, a few mG \citep[][]{gi06}.

Recently, \cite{hou09} showed  how the signal integration through the thickness of the cloud 
and the area of the telescope beam affects on the measured angular dispersion and 
apply their results to OMC-1. 
Based on their estimated number (N=21) of the independent turbulent cells contained 
within the column probed by the telescope beam, 
they found that a correction factor of $1/\sqrt{N} \sim0.2$ is applicable to OMC-1. 
In our case, although the area of the telescope beam is negligibly small due to 
the point sources, the thickness of the cloud should be taken into account and 
the correction factor should be somewhat smaller than $\sim0.5$. 
If we assume that the effect of the cloud thickness is similar to that of OMC-1, 
we obtain $N\sim11$, suggesting a factor of $\sim0.3$. 
Adopting this factor of $\sim0.3$, the above estimated values are reduced 
by a factor of $\sim0.6$ and $B_{\parallel}\sim100$ may be appropriate 
for the ambient region of the Serpens cloud core, except the SW corner.

Here, we roughly derive the mass to magnetic flux ratio $M_{\rm cloud}/\Psi$
using our estimated value of  $B \sim$100 $\mu$G,  
and compare it with the critical value for a magnetic stability of the cloud, 
 ($M_{\rm cloud}/\Psi$)$_{\rm critical}=(4\pi^2 G)^{-1/2}$ \citep{nak78}.
With a formula  $M_{\rm cloud}/\Psi = (\pi R^2 \mu m_{\rm H} N)/(\pi R^2 B)=\mu m_{\rm H}N/B$ 
and the H$_2$ column density $N\sim 6 \times10^{22}$ cm$^{-2}$ 
where we estimated $B$,  
we derive $M_{\rm cloud}/\Psi \sim3.8 \times(M_{\rm cloud}/\Psi)_{\rm critical}$, 
where $R$ is a radius of the cloud and $m_{\rm H} $ is the mass of a hydrogen atom.  
Although this derived value is slightly larger than the critical value, 
$M_{\rm cloud}/\Psi$ could be much larger in the inner region of the cloud core
because the column density of the inner region is much higher than those 
where we estimated $B$, but the magnetic field may be slightly larger than that 
we estimated in the ambient region, judged from the slowly curved shape of the magnetic field. 
We note that the adopted strength of the magnetic field is that estimated for 
the projection of the magnetic field in the plane of the sky, suggesting a slightly smaller 
$M_{\rm cloud}/\Psi$ than the estimated one.
These imply that  the ambient region is marginally supercritical, 
while the inner region is supercritical.  
This situation is considered to be quite consistent with the hourglass shape 
of the magnetic field and with the cluster formation within the sub-clumps.

It is interesting to examine whether the magnetic field can maintain the outflow collimation 
along the magnetic field in the ambient region of the sub-clumps, 
i.e., whether the magnetic field can guide the outflows. 
The magnetic pressure, $P_{B}=B^2/8\pi$, is calculated to be 
$\sim$ 4 $\times $10$^{-10}$  dyn, adopting $B\sim100 \mu$G. 
Assuming the average density and velocity width due to turbulence for the outflow to be $3 \times 10^3$ cm$^{-3}$, 
which would be consistent with the optically thin condition of the high velocity gas \citep{wh95}, 
and 3 km s$^{-1}$, which is larger than the C$^{18}$O velocity width by a factor of $\sim$1.5--2.0, 
we obtain the turbulent pressure, 
$P_{\rm turb}=\rho \sigma_{\rm turb}^2$, of $\sim 2 \times $10$^{-10}$  dyn. 
Taking into account the fact that the adopted strength of the magnetic field is 
that estimated for the projection and that $P_{B}$ is proportional to $B^2$,
these estimates imply that the magnetic field can guide the outflows in the ambient region of the Serpen cloud core.

\subsection{Comparison with outflow-driven turbulence model for cluster formation}

From our analysis, the magnetic field seems to be important in considering 
the cloud stability that is related to star formation or cluster formation 
and the feed back from the star formation activity, such as outflows. 

The hourglass-shaped magnetic field suggests that the Serpens cloud core 
first contracted along the straight magnetic field to be a filament or elongated cloud, 
which is perpendicular to the magnetic field, and that then the central part contracted 
cross the magnetic field due to the high density in the central region of the cloud core. 
This situation is very similar to the contraction of the low-mass core that is penetrated 
by the uniform magnetic field \cite[e,g.,][]{gi06, kan09}.
In addition, there might exist the cloud rotation, of which axis agrees with that of 
the hourglass-shaped magnetic field.
It was reported that many small-scale outflows spread to or penetrate the NW and SW 
sub-clupms\citep[e.g.,][]{her97,ho99,da99,zi99}, 
and the ambient, larger-scale outflows (filaments) seem to run along the magnetic field 
as shown above \citep{da99,nar02}.  
Moreover, it is possible that the blue-shifted $^{13}$CO ($1-0$) features just outside CLS 1, 
which correspond to the red-shifted CO ($2-1$) outflows, are inflows from the ambient 
to the central part of the SE sub-clump. 
Considering these altogether, we may have to take into account the magnetic field, 
outflows, inflows, cloud rotation, and contraction as well as the turbulence of the molecular gas 
in the cluster formation process of the Serpens cloud core (see Figure \ref{f10}).

The structures mentioned above seem to be in good agreement with the outflow-driven 
turbulence modelof \cite{li06} and \cite{nakam07} who performed 3D MHD simulation of 
cluster formation takinginto accout the effect of protostellar outflows. 
They demonstrated that protostellar outflows can generate supersonic turbulence 
in pc-scale cluster forming clumps like the Serpens cloud core.
One of the important characteristics of outflow-driven turbulence is 
that gravitational infall motions almost balance the outward motions driven by outflows, 
creating very complicated density and velocity structure \citep[see e.g., Figure 4 of][]{nakam07}.  
The resulting quasi-equilibrium state can be maintained through active star formation 
in the central dense region. 
In the presence of relatively strong magnetic field, both outflow and inflow motions 
in the less dense envelope tend to be guided by large scale ordered magnetic field lines.   
As a result, filamentary strucutures that are roughly converging toward the central dense region
appear in the envelope, whereas the density structure tends to be more complicated in the central 
dense region where self-gravity and turbulence may dominate over the magnetic field. 
Infall motions detected by $^{13}$CO ($1-0$) in the Serpens core may correspond to such filamentary structures created by gravitational infall.

To clarify how the outflows and magnetic field affect the dynamical state of the cloud, 
we assess the force balance in the cloud, following  \cite{mau09}.
To prevent the global gravitational contraction, the following 
pressure gradient is needed to achieve the hydrostatic equilibrium:
\begin{equation}
{dP_{\rm grav} \over dr} \simeq -G{M(r) \rho (r) \over r^2} \left(1-\alpha^{-2}\right) \ ,
\end{equation}
where $M(r)$ is the mass contained within the radius $r$ and we assume that the cloud is spherical.
The effect of magnetic field is taken into account by the factor $(1-\alpha^{-2})$
and $\alpha$ is the mass-to-magnetic flux ratio normalized to the
critical value  and is approximated as 
\begin{equation}
\alpha \simeq {2\pi G^{1/2} M/\pi r^2 \over B}
\end{equation}
 \citep[e.g.,][]{nak98}.

Assuming the density profile of $\rho \propto r^{-2}$, the pressure 
needed to support the cloud against the gravity is estimated to be
\begin{equation}
P_{\rm grav} \simeq {GM(R)^2 \over 8\pi R^4} \left(1-\alpha^{-2}\right) \ .
\end{equation}
The force needed to balance the gravitational force is thus evaluated to be 
\begin{equation}
F_{\rm grav} \simeq 4 \pi R^2 P_{\rm grav} (R) = {GM(R)^2 \over 2 R^2} \left(1-\alpha^{-2}\right) \ .
\end{equation}
Adopting $M(R)= 210 M_\odot$, $R=0.46$ pc \citep{ol02}, 
$B=100 \mu$G and $\alpha = 3.8$, 
$F_{\rm grav}$ can be estimated to be $ \sim 4.3 \times 10^{-4} M_\odot$ km s$^{-1}$ yr$^{-1}$. 
The moderately strong magnetic field of $\alpha = 3.8$ can 
reduce the gravitational force by $\sim$7\%.
We note that we rescaled the cloud mass and radius derived from \cite{ol02}  
by assumingthe distance to the cloud of 260 pc.  
Hereafter, we also use other values rescaled for this distance.  

On the basis of the CO ($J=2-1$) observations, \cite{da99} 
detected many powerful CO outflows in this cloud, and derived 
the physical properties of the outflows. From their analysis, 
we can evaluate the total force exerted by the outflows in this region
as 
\begin{equation}
F_{\rm outflow} \simeq {p_{\rm outflow} \over t_{\rm dyn}} \sim 
{8.7\textendash17.5M_\odot {\rm km \ s}^{-1} \over 2.5 \times 10^4 {\rm yr}} \sim
(3.4\textendash7.0) \times 10^{-4} 
M_\odot \ {\rm km \ s^{-1} \ yr^{-1}}  \ 
\end{equation} 
where $p_{\rm outflow}$ is the total outflow momentum, and $t_{\rm dyn}$ is the 
representative dynamical time of the outflows.
The force due to the outflows, $F_{\rm outflow}$, is comparable to or somewhat larger 
than the force needed to stop the global gravitational collapse, $F_{\rm grav}$, 
suggesting that the outflows play a crucial role in the cloud dynamics.
This result, however, apparently contradicts that of \cite{ol02} who
suggested that the cloud may be undergoing a global contraction, although 
the further justification is needed to confirm their interpretation.
This apparent inconsistency may come from our assumption of the spherical cloud.
Since the relatively strong magnetic field associated with the cloud can 
guide the large scale outflow motions along the global magnetic field as discussed 
in the previous subsection, the force exerted by the outflows is expected to be 
weak along the cross-field direction. 
As a result, the cloud may be able to contract along the cross-field direction.
For the Serpens core, both the magnetic field and the outflows are likely to control the cloud dynamics.

The outflows are also expected to be the major source for generating 
supersonic turbulence in the Serpens core.
From the physical quantities of the outflows measured by \cite{da99},
we can evaluate the total energy injection rate due to the outflows in this region
as
\begin{equation}
{dE_{\rm outflow} \over dt} \simeq {E_{\rm outflow} \over t_{\rm dyn}} \sim 
{(12.7\textendash48.3) \ {\rm J} \over 2.5\times 10^4 \ {\rm yr}} \sim (0.5\textendash2) L_\odot   \ .
\end{equation}
where $E_{\rm outflow}$ is the total outflow energy.  
The energy dissipation rate of supersonic turbulence is obtained by \citet{mac99} as
\begin{equation}
{dE_{\rm turb} \over dt} = f {1/2 M \Delta V^2 \over \lambda_d / \Delta V}
\end{equation}
where $f (=0.34)$ is the non-dimensional constant determined from the numerical simulations,
and $M$ is the cloud mass, and $\Delta V$ is the 1D FWHM velocity width.
The driving scale of the turbulence $\lambda_d$ is estimated to be $\lambda_d \sim 0.4$ pc
for the outflow-driven turbulence \citep{ma07, nakam07}.
The energy dissipation rate of the turbulence can be estimated to be 
 $\sim 0.12 L_\sun$, 
where the FWHM velocity width of about 2 km s$^{-1}$ is adopted \citep{ol02}.  
This energy dissipation rate is somewhat smaller than
the outflow energy input rate.
In the Serpens cloud core, the relatively strong magnetic field tends to guide the 
outflows and therefore the significant amount of the outflow energy might escape 
away from the cloud along the magnetic field, as inferred from the magnetic field 
and outflow structures discussed above. In any case, the outflows seem to have sufficient
energy to power supersonic turbulence in this region
and the magnetic field seems to play an important role in the escape of the outflow energy 
from the cloud.
These characteristics appear to be in agreement with the the outflow-driven turbulence model 
for cluster formation, and imply the importance of the magnetic field for 
the continuous star formation  in the central region of the Serpens cloud core 
under the condition where the outflow energy injection rate is high.
The Serpens cloud core is expected to be one of the good examples of 
the outflow-driven turbulence model for cluster formation.

\subsection{Summary}

We have conducted deep and wide ($\sim$7\farcm7 $\times$ 7\farcm7) $JHK_{\rm s}$ 
imaging polarimetry of the Serpens cloud core.    The main findings are as follows:

1. The central part of the infrared reflection nebula is illuminated mainly by two sources; 
the north by SVS 2 (SRN)  and the south by SVS 20 with two centrosymmetric patterns.
The characteristics of the nebula are consistent with those reported 
in the previous infrared polarimetric works.
 Detailed inspection enabled us to find  24 YSOs associated with IR nebulae, 
in addition to SVS 2 and SVS 20.   

2. Polarization of NIR point sources was measured and those sources, except YSOs, 
have an upper limit of polarization degree similar to that of the nearby star forming regions. 
It is consistent with the dichroic origin, i.e., the polarization vectors of the near-IR point sources 
could indicate the direction of the averaged local magnetic field.

3. The polarization vectors suggest a clear hourglass shape. 
We have made a model fitting of this shape with a parabolic function and found that 
the symmetry axis ($\theta_{\rm PA} \sim$70\degr)  of the hourglass magnetic field 
is nearly perpendicular to the elongation ($\sim150 \degr$) of the bright parts of C$^{18}$O ($J=1-0$)  
or submillimeter continuum emissions, i.e., the alignment direction of NW and SE sub-clumps.  
The submillimeter continuum filaments and CO outflow lobes, which protrude 
from these sub-clumps, seems to run along the best-fit magnetic field in the ambient region
and some $^{13}$CO velocity features also seem to be along the magnetic field.

4. The evaluation of the magnetic field strength has been done with the CF method 
 toward the ambient area of the Serpens cloud core, taking into account the recent study 
on the signal integration effect for the dispersion component of the magnetic field.  
The mass to magnetic flux ratio was estimated with the evaluated magnetic field strength 
of $\sim100\mu$G and the parameters of the previous C$^{18}$O ($J=1-0$) observations, 
and found to be slightly larger than the critical value of magnetic instability 
in the  the ambient area.  
This suggests a possibility that the central region is magnetically unstable, 
which is consistent with the fact that star formation is actively taking place in the central region.
We estimated the magnetic pressure and the turbulent pressure of the outflow 
using the evaluated magnetic field strength and possible turbulent parameters, 
and found that  the magnetic pressure could be high enough to  guide the outflows 
in the ambient region.

5. The bright part of C$^{18}$O ($J=1-0$),  submillimeter continuum cores as well as 
many class 0/I objects are located just toward the constriction region of 
the hourglass-shaped magnetic field.
These suggest that the Serpens cloud core first contracted along the magnetic field to be 
an elongated cloud and that then the central part contracted cross the magnetic field 
due to the high density in the central region of the cloud core. 

6.  Comparisons of the best-fit magnetic field with the previous observations of molecular gas 
and large-scale outflows suggest  a possibility that the cloud dynamics is controlled by the magnetic field, 
protostellar outflows and gravitational inflows. 
In addition, the outflow energy injection rate appears to be  the same as or larger than 
the dissipation rate of the turbulent energy in this cloud, 
indicating that the outflows are the main source of turbulence 
and that the magnetic field plays an important role both in 
allowing the outflow energy to escape from the central region of the cloud core 
and enabling the gravitational inflows from the ambient region to the central region.
These characteristics appear to be in good agreement with the outflow-driven turbulence model 
for cluster formation and imply the importance of the magnetic field to continuous star formation 
in the center region.

\acknowledgments

We thank Chris Davis for kindly providing us the 850 $\mu$m continuum data. 
This work was supported by Grant-in-Aid for Scientific Research (20403003, 19540242, 
and 19204018) from the Ministry of Education, Culture, Sports, Science and Technology.






\appendix

\section{Identification of YSOs with NIR reflection nebulae toward the central region 
of the Serpens cloud core}

\begin{figure}
\plotone{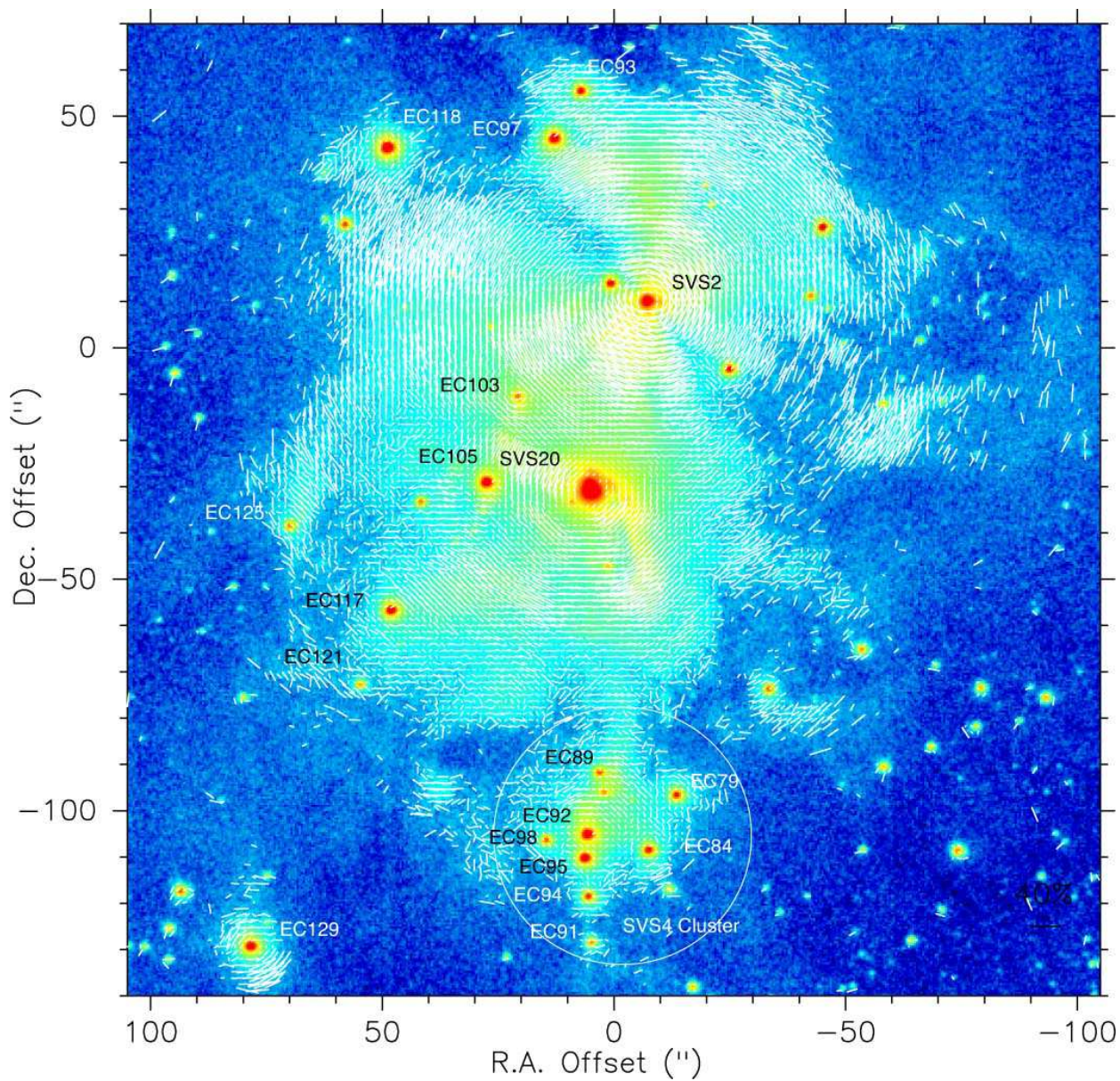}
\caption{$K$s polarization vector map of the central region of the Serpens cloud core, 
superposed on the total intensity images in the logarithmic scale.
The vectors were made by 3$\times$3 pixel binning. 
The area of the image is $220\arcsec \times 220 \arcsec.$  
The reference position of offset is $(\alpha, {\mathit \delta})_ {\rm J2000}$ 
=(18:29:57.39, +01:14:36.2).  North is at the top and east is to the left.  
40\% vector is shown near the bottom right corner.  
The young stellar objects with reflection nebulae are marked.
The SVS 4 cluster is shown by the circle. \label{f3}}
\end{figure}

\begin{figure}
\plotone{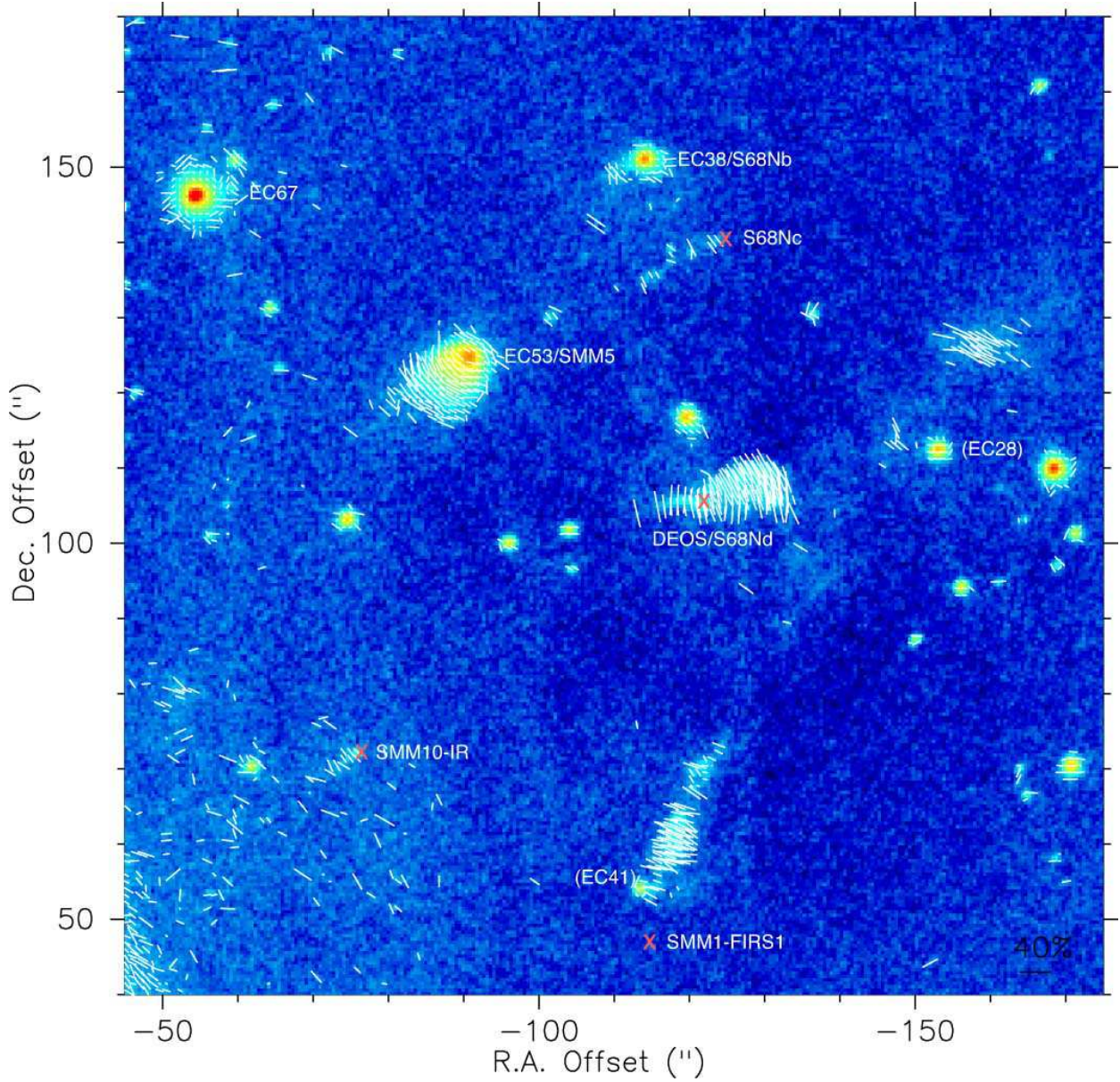}
\caption{$K$s polarization vector map of the NW sub-core region of 
the Serpens cloud core, superposed on the total intensity images 
in the logarithmic scale.  
The vectors were shown every 2 pixels.
The area of the image is $130\arcsec \times 130 \arcsec.$  
The reference position of offset is $(\alpha, {\mathit \delta})_ {\rm J2000}$ 
=(18:29:57.39, +01:14:36.2).  North is at the top and east is to the left.  
40\% vector is shown near the bottom right corner.   
The young stellar objects with reflection nebulae are marked, but EC28 and EC41 
are not identified as young stellar objects with reflection nebulae. \label{f4}}
\end{figure}

\begin{figure}
\begin{center}
\includegraphics[scale=.90]{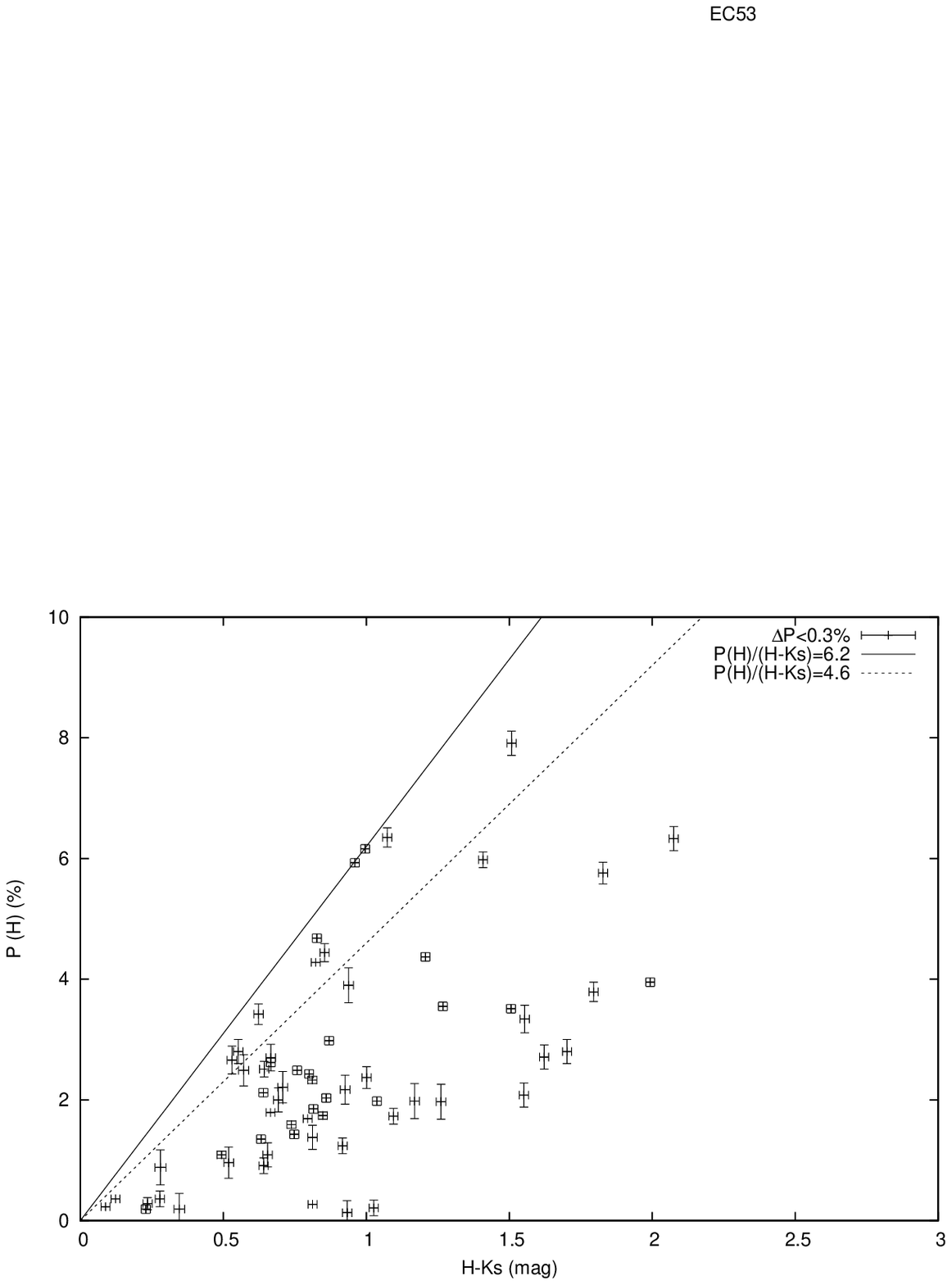}\\
\includegraphics[scale=.90]{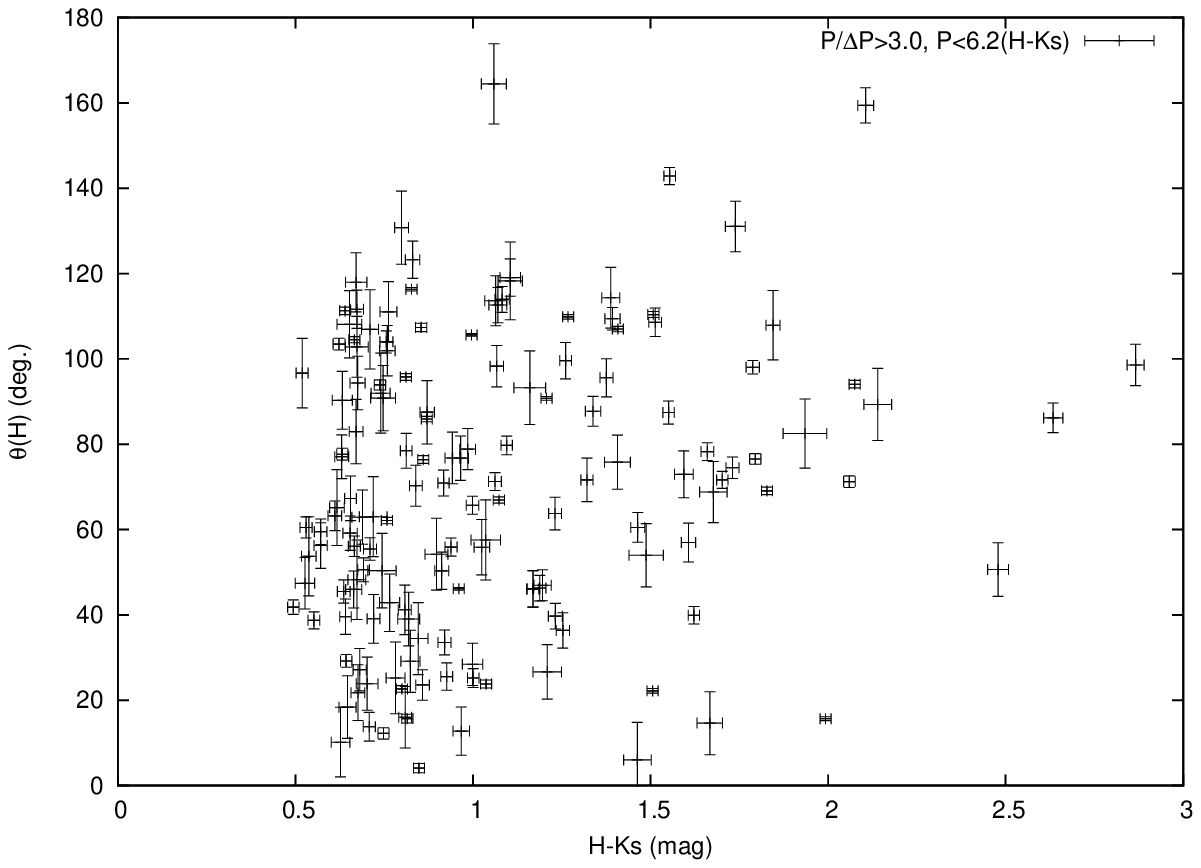}
\end{center}
\caption{Top: polarization degrees at $H$ versus $H-K$s color for sources having the 
$H$-band polarization errors of $<0.3\%$.  Bottom: polarization P.A.s 
at $H$ versus $H-K$ color for sources having  $P/{\mathit \Delta} P > 3$ and 
$P <  6.2 (H-Ks)$.\label{f5}}

\end{figure}

\begin{figure}
\epsscale{.90}
\plotone{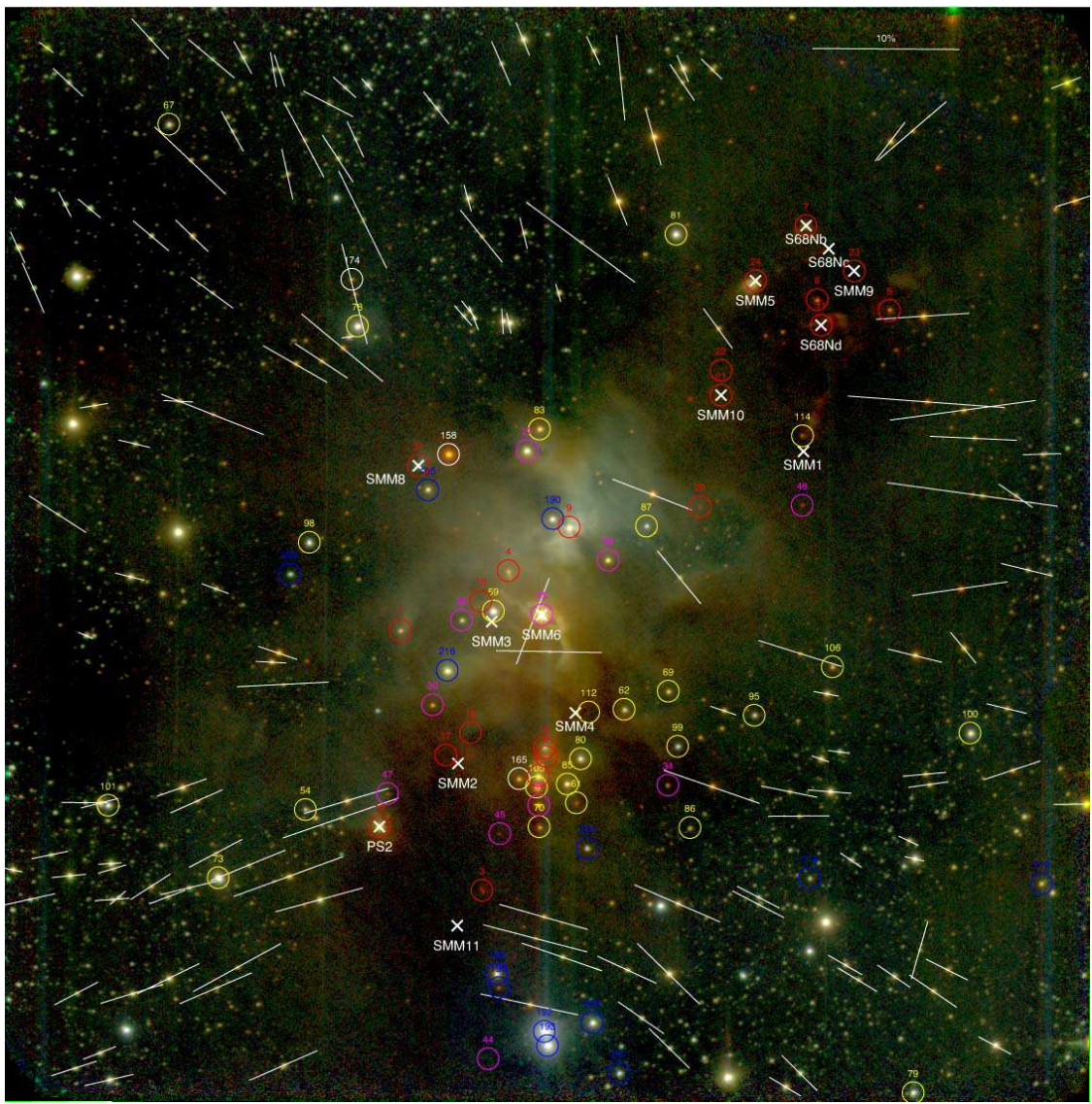}
\caption{$H$-band polarization vectors (white bars) for point sources having 
$P/{\mathit \Delta} P > 3$ and $P < 6.2 (H-Ks)$, 
superposed on a $JHK$s composite color image in the logarithmic scale. 
10\% vector is shown near the upper right corner.  
The area of  the image ($\sim 8\farcm4 \times 8\farcm5$) made 
by averaging the dithered frames is slightly larger than that of the camera array  
($\sim 7\farcm7 \times 7\farcm7$).  North is at the top and east is to the left.  
YSOs identified by the Spitzer space telescope are marked by circles 
(Class 0/I; red, Flat-spectrum; magenta, Class II; yellow, Transition disk; blue, 
Class III; white) with ID numbers \citep[Tables 3 and 4 of][]{wi07}. 
Submillimeter continuum peaks are show by crosses with names and 
their coordinates come from the Spitzer photometry of  \cite{wi07}, 
except SMM2 and SMM11 \citep{da99}. \label{f6}}
\end{figure}

\begin{figure}
\epsscale{1.0}
\plotone{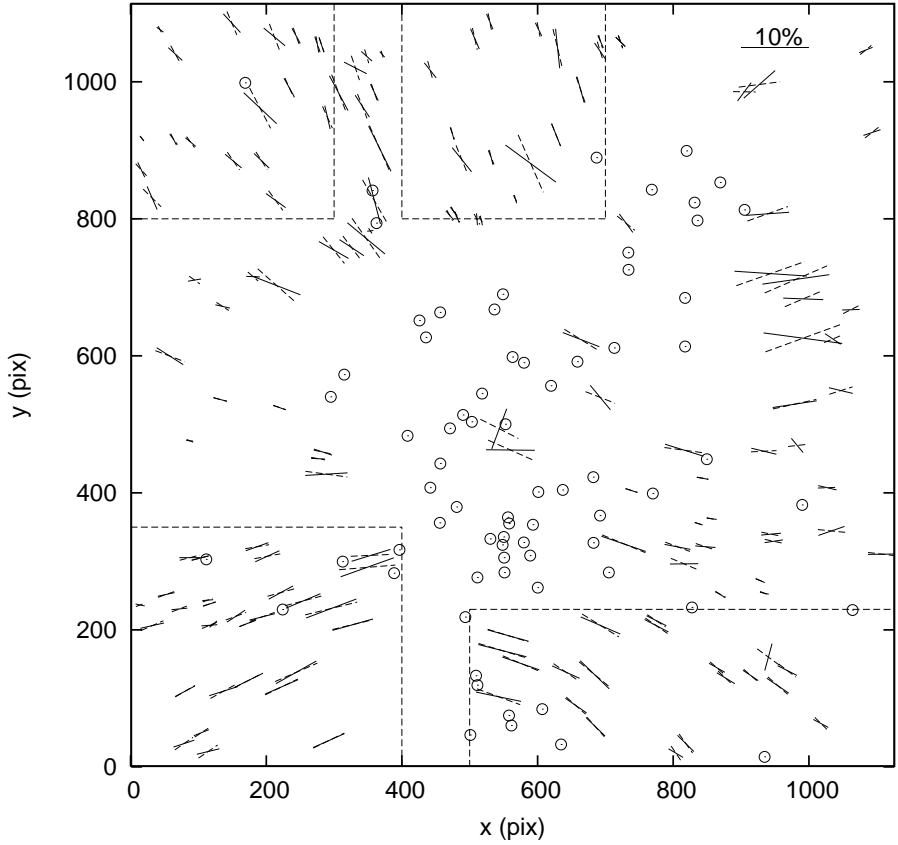}
\caption{$H$-band measured polarization vectors (solid bars) for point sources 
having $P/{\mathit \Delta} P > 3$ and $P < 6.2 (H-K{\rm s})$ and their best-fit polarization 
(magnetic field) vectors (dashed bars) with a parabolic function 
($x=g+gCy^2$), where the $x$ is the distance from the parabolic magnetic field axis 
of symmetry having the coefficient of $y^2$ of $C=(7.00\pm 0.39)\times 10^{-6}$ pixel$^{-2}$
and the symmetry center of the magnetic field is 
$(x, y)_{\rm center}=(646.0\pm11.3, 384.0\pm9.1$), which corresponds to 
$(\alpha, \delta)_ {\rm J2000}$ =(18:29:54.9, +01:13:13), 
on the axis of symmetry that is tilted  at $\theta_{\rm PA}=70.01\degr \pm0.77\degr$.  
$10\%$ vector is shown near the upper right corner.  
The area of  the figure is $\sim 8\farcm4 \times 8\farcm5$, nearly same as that of Fig. \ref{f6}.  
YSOs identified by the Spitzer space telescope are marked by circles, and these YSOs are
not used for the fitting.
In four areas enclosed by dashed lines,  additional fitting 
with one-parameter ($C$) was done (see text).  
\label{f7}}
\end{figure}

\begin{figure}
\epsscale{.90}
\plotone{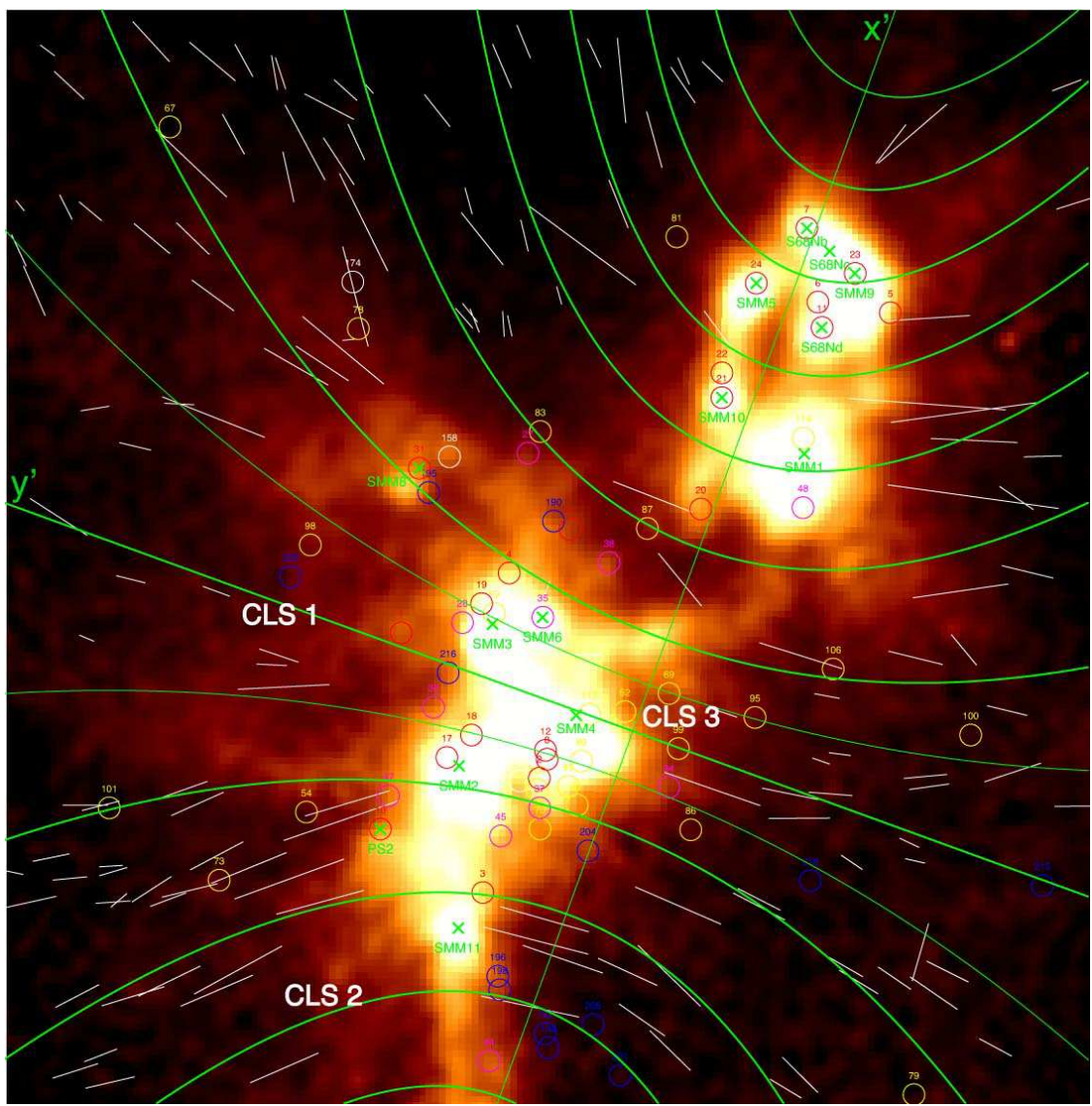}
\caption{$H$-band polarization vectors (white lines) for point sources having 
$P/{\mathit \Delta} P > 3$ and $P < 6.2(H-K$s), and their best-fit magnetic field  
(green thick curved/ straight lines and thin curved lines), superposed on SCUBA 850 $\micron$ 
continuum image \citep[kindly provided by C.J. Davis, see Figure 1 of][]{da99}. 
Note that these lines do not present lines of magnetic force, just the direction of the magnetic field.
The horizontal axis (green thin straight line), which is perpendicular to the parabolic 
magnetic field axis (green thick straight line), is also shown. 
The area of  the image is $\sim 8\farcm4 \times 8\farcm5$, same as that of Fig. \ref{f6}.  
North is at the top and east is to the left.  
The YSOs identified by the Spitzer space telescope and submillimeter continuum peaks 
are indicated the same as in Figure \ref{f6}.\label{f8}}
\end{figure}

\begin{figure}
\epsscale{.80}
\plotone{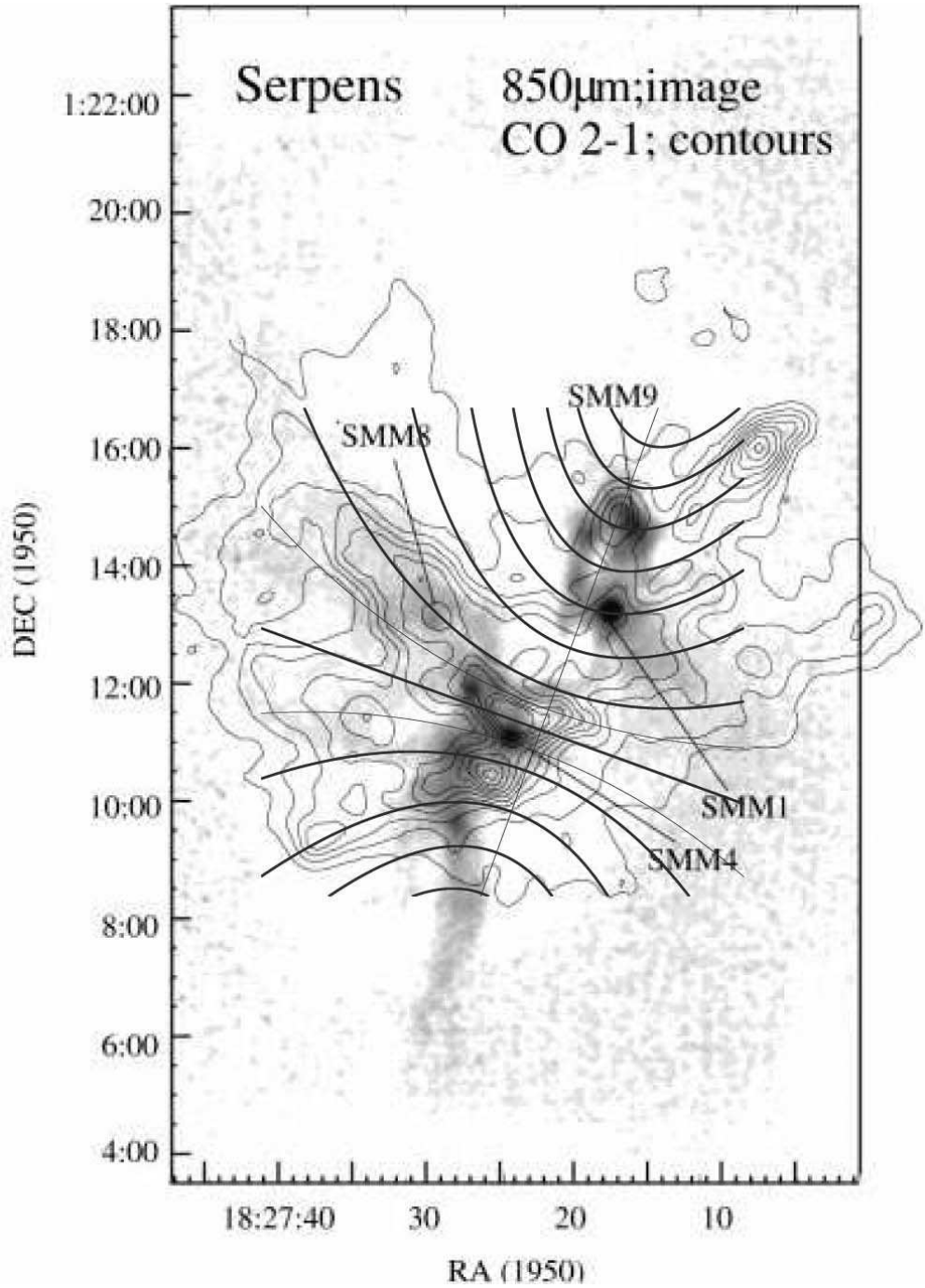}
\caption{The best-fit magnetic field (the same as Figure \ref{f8}),
 superposed on  the CO $J=2-1$ contour plot of $V_{\rm LSR}$=2--16 km s$^{-1}$ 
 and 850 $\micron$ continuum image \citep[Figure 2 of][]{da99}.  
\label{f9}}
\end{figure}

\begin{figure}
\plotone{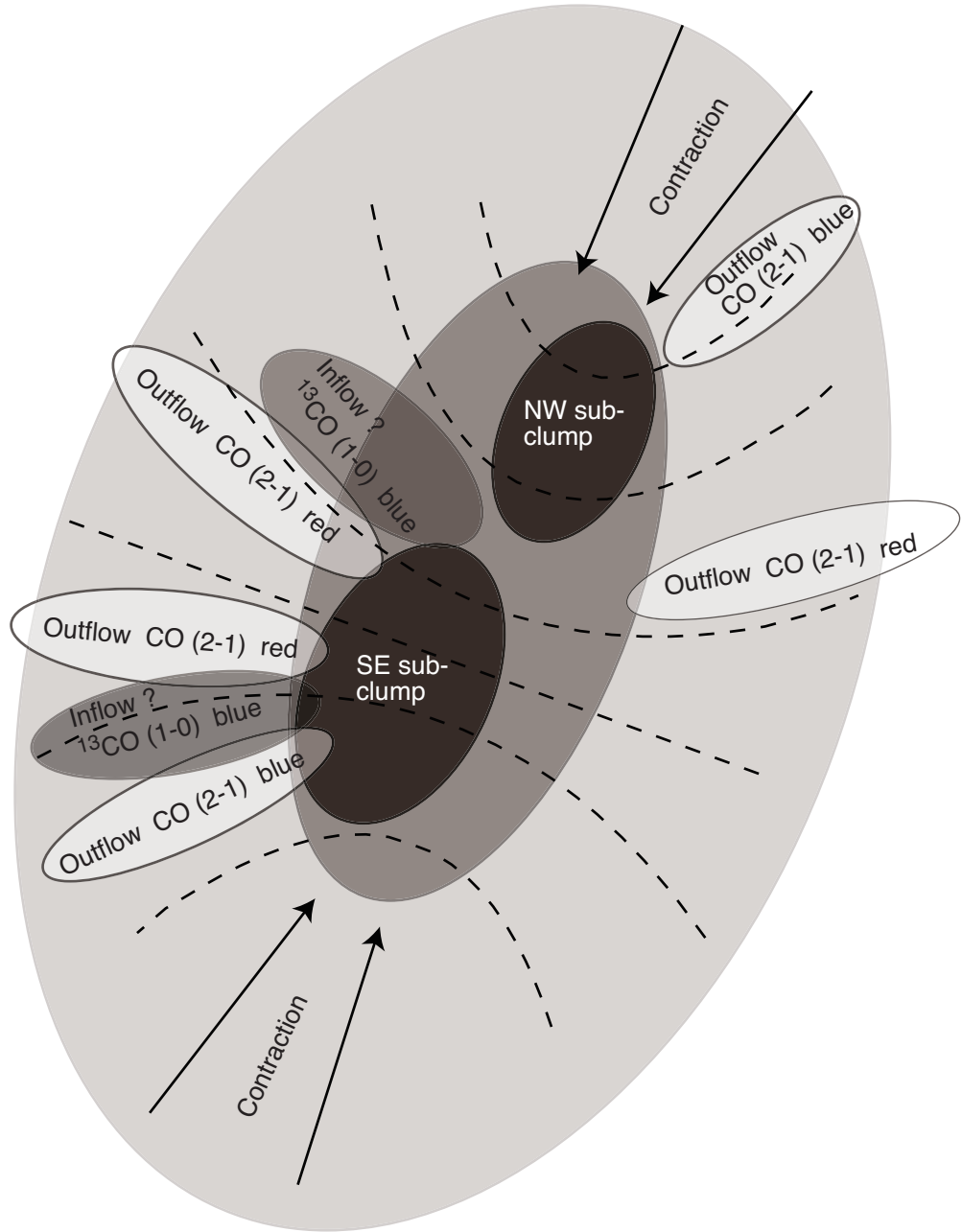}
\caption{The schematic drawing of the Serpens cloud core. The pinched or hourglass-shaped magnetic field pattern suggests that the magnetic field lines are dragged along with the contracting gas toward the center of the cluster.\label{f10}}
\end{figure}

\begin{figure}
\plotone{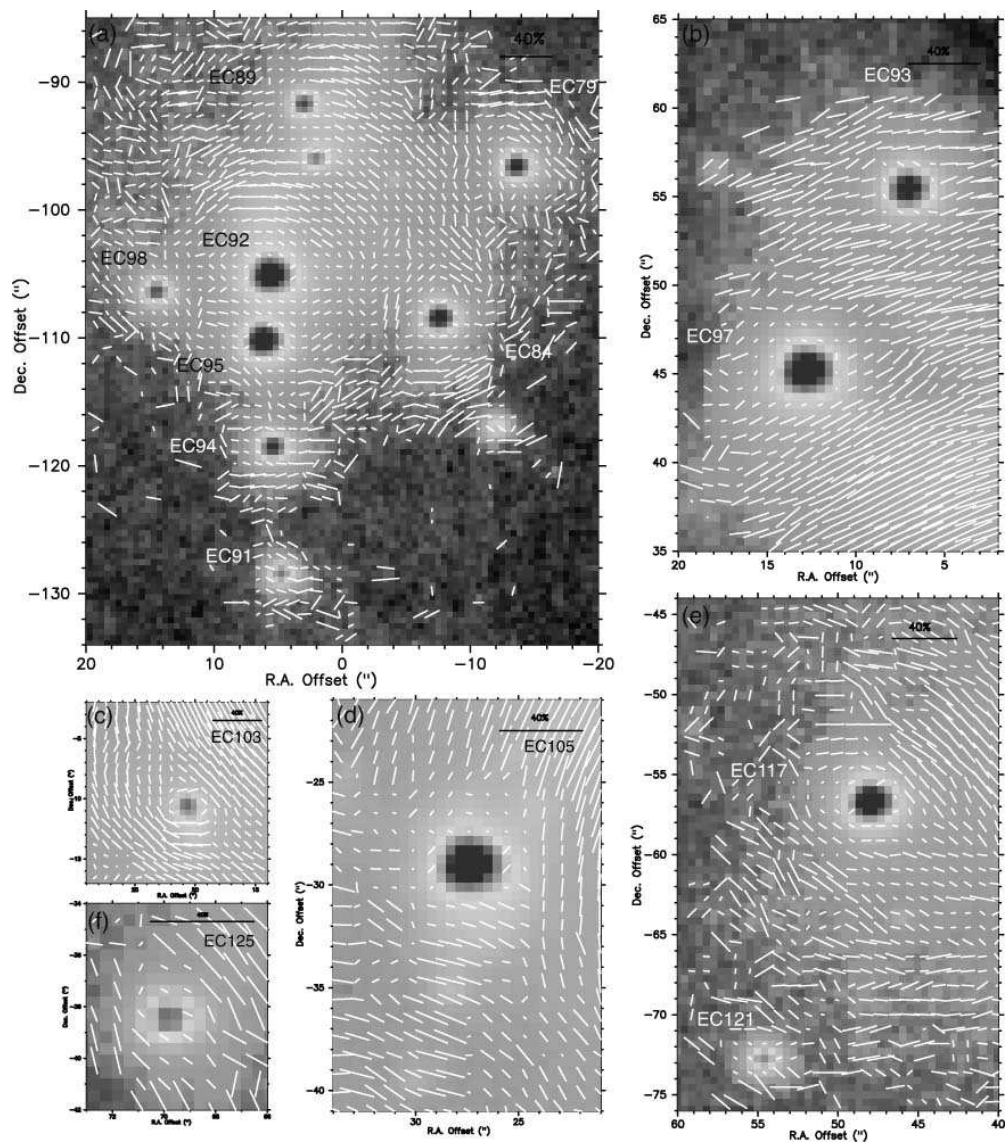}

\caption{$K$s polarization vector maps of the nebulosities associated with YSOs 
in the SVS 4 cluster and the central region of the Serpens cloud core, superposed 
on the total intensity images in the logarithmic scale.  
The vectors were shown every 2 pixels.
The reference position of offset is $(\alpha, {\mathit \delta})_ {\rm J2000}$ 
=(18:29:57.39, +01:14:36.2).  North is at the top and east is to the left.  
40\% vector is shown at each panel.   
YSOs with reflection nebulae are marked.  \label{f11}}
\end{figure}

Except isolated YSOs that are on the periphery of  the central region, 
it is not easy to examine whether YSOs have reflection nebulae locally illuminated 
by themselves only with the binned map of the $K$ s polarization vectors (Figure \ref{f3}), 
due to the contamination of light from the strong sources or nearby sources.  
We constructed the highest resolution maps without binning
for the sources suffering from the contamination of SVS 2, SVS 20 and 
the members of the SVS 4 cluster, and tried to identify which sources are associated 
with reflection nebulae. 
In Figure  \ref{f11}, we show the polarization vector maps only for YSOs that we have 
identified as those having reflection nebulae.
For EC 94, EC 98, and EC 121, although the vector map quality/resolution is not always 
good enough for the robust identification, we concluded, taking account  the weak emission around these sources, that they probably have reflection nebulae. 




\clearpage



\begin{deluxetable}{lccccl}
\tabletypesize{\scriptsize}
\tablecaption{YSOs with NIR nebulae  toward the central region.\label{tbl_a}}
\tablehead{
\colhead{YSO  Name\tablenotemark{a} } & \colhead{$K$s} &  \colhead{Source}  & \colhead{SSTc2dJ} & \colhead{ISO} & \colhead{Other Name\tablenotemark{a}}  \\
\colhead{(Spitzer ID\tablenotemark{b} )} & \colhead{(mag)\tablenotemark{b}} & \colhead{Class\tablenotemark{b} } & \colhead{ID\tablenotemark{c}} & \colhead{ID\tablenotemark{d}} & \colhead{}
}
\startdata

EC79 (80)\tablenotemark{e} & 11.5 & 2 & 18295655+0112595 & 304 & GCNM84/STGM12  \\
SVS2 (9)   &   9.3 & 0/1 & 18295687+0114465 & 307 & EC82/GCNM87/CK3/STGM22 \\
EC84 (85)\tablenotemark{e} & 11.1  & 2  & 18295696+0112477 & 309  & GCNM90/STGM11  \\
EC89 (12)\tablenotemark{e} & 12.0 & 0/1 & 18295766+0113046 & (312) & GCNM97/STGM13  \\
SVS20S/N (35) & 7.1 & FS & 18295772+0114057 & 314 & EC90/GCNM98/CK1/STGM18 \\
EC91 (70)\tablenotemark{e}  &  13.0 & 2 & 18295780+0112279 & 320 & GCNM101 \\
EC92 (2)\tablenotemark{e} & 10.5  & 0/1  & 18295783+0112514 & (317) & GCNM104  \\
EC93 (83) &  10.8 & 2 & 18295780+0115318 & 319 & GCNM100/STGM25/CK12 \\
EC94 (37) \tablenotemark{e}  &  11.7   & FS & 18295784+0112378 & 318 & GCNM102 \\
EC95 (105)\tablenotemark{e} & 10.0  & 2  & 18295789+0112462  & (317) & GCNM103  \\
EC97 (27) & 9.9 & FS & 18295819+0115218 & 321 & GCNM106/CK4/STGM24 \\
EC98 (165)\tablenotemark{e} &  12.3 & TD & 18295844+0112501 & 322 & GCNM110 \\
EC103 (4) & 11.8 & 0/1 & 18295877+0114262 & 326 & GCNM112/STGM20/K2\_5 \\
EC105 (59) & 9.5 & 2 & 18295923+0114077 & 328 & GCNM119/CK8/STGM19/K2\_6 \\
EC117 (216) & 10.1 & 3 & 18300065+0113402 & 338 & GCNM135/CK6 \\
EC118 (158) & 9.0 & TD & 18300061+0115204 & 337 & GCNM136/CK2 \\
EC121 (30)   & 13.2 & FS & 18300109+0113244 & 341 & GCNM142 \\
EC125 (1)   & 13.4 & 0/1 & 18300208+0113589 & 345 & GCNM154/CK7/STGM16 \\
EC129 (10) &  9.9  & 0/1 & 18300273+0112282 & 347 & GCNM160/STGM10 \\

\enddata

\tablenotetext{a}{From the names referred in Table 1 of \cite{ei08}}
\tablenotetext{b}{From \cite{wi07}}
\tablenotetext{c}{From \cite{ha07}}
\tablenotetext{d}{From \cite{ka04}}

\tablenotetext{e}{SVS-4 cluster}
\end{deluxetable}

\clearpage

\begin{deluxetable}{lccccl}
\tabletypesize{\scriptsize}					
\tablecaption{YSOs with NIR nebulae  toward the NW region.\label{tbl_b}}
\tablehead{
\colhead{YSO  Name\tablenotemark{a} } & \colhead{$K$s} &  \colhead{Source}  & \colhead{SSTc2dJ} & \colhead{ISO} & \colhead{Other Name\tablenotemark{a}}  \\
\colhead{(Spitzer ID\tablenotemark{b} )} & \colhead{(mag)\tablenotemark{b}} & \colhead{Class\tablenotemark{b} } & \colhead{ID\tablenotemark{c}} & \colhead{ID\tablenotemark{d}} & \colhead{}
}
\startdata

DEOS/S68Nd\tablenotemark{e} (11) & 14.9 & 0/1 &18294913+0116198 & 250 & knot c\tablenotemark{f}/K4\_5/WMW11  \\
S68Nc\tablenotemark{e}  & \nodata & (0/1) & \nodata & \nodata & subknot a3\tablenotemark{e}/knot a\tablenotemark{f} \\
SMM1-FIRS1 & \nodata & \nodata  & 18294963+0115219 & 258a  & GCNM23/WMW114/VLA7  \\
EC53/SMM5 (24) & 11.3\tablenotemark{c} & 0/1 & 18295114+0116406 & 265 & STGM27/WMW24  \\
EC67 (81) & 9.6 & 2 &18295359+0117018 & 283 & GCNM60/STGM29/WMW81  \\
EC38/S68Nb\tablenotemark{e} (7) & 12.7  & 0/1  & 18294957+0117060 & 254 & WMW7  \\
SMM10-IR (21) & 17.7\tablenotemark{c}   & 0/1  & 18295219+0115478  & 270 & WMW21  \\

\enddata

\tablenotetext{a}{From the names referred in Table 1 of \cite{ei08}}
\tablenotetext{b}{From \cite{wi07}}
\tablenotetext{c}{From \cite{ha07}}
\tablenotetext{d}{From \cite{ka04}}
\tablenotetext{e}{From \cite{wil00}}
\tablenotetext{f}{From \cite{da99}}
\tablenotetext{g}{From \cite{ho99}}
\end{deluxetable}






\end{document}